\documentclass{aa}  

\usepackage{graphicx}
\usepackage{txfonts}
\usepackage{textcomp}
\usepackage{subfig}
\usepackage[usenames,dvipsnames]{color}
\usepackage[normalem]{ulem}
\begin{document}

   \title{Implications of the correlation between bulge-to-total baryonic mass ratio and the number of satellites for SAGA galaxies}

   \author{A. Vudragovi\'c
          \inst{1},
          I. Petra\v s\inst{2}, M. Jovanovi\'c\inst{1}, S. Kne\v zevi\' c\inst{1} \and S. Samurovi\'c\inst{1}
          }

   \institute{Astronomical Observatory of Belgrade,
              Volgina 7, 11060 Belgrade, Serbia\\
              \email{ana@aob.rs}
         \and
             Faculty of Mathematics, Studentski trg 16, 11000 Belgrade, Serbia
             }

   \date{Received September 15, 1996; accepted March 16, 1997}

 
  \abstract
   {We searched for correlations between the number of satellites and fundamental galactic properties for the Milky Way-like host galaxies in order to better understand their diverse satellite populations. We specifically aim to understand why galaxies that are very similar in stellar mass content, star formation rate, and local environment have very different numbers of satellites.  
   }
   {Deep and extensive spectroscopic observations are needed to characterize the complete satellite luminosity function beyond the Local Group. One such endeavor is an ongoing  Satellites of Galactic Analogs (SAGA) spectroscopic survey that has completed spectroscopic observations of 36 Milky Way-like galaxies within their virial radii down to the luminosity of Leo I dwarf galaxy. We correlated the number of satellites of SAGA galaxies with several fundamental properties of their hosts ---including total specific angular momentum, which is considered to be well preserved throughout galaxy lifetime--- in an attempt to identify the main driver of their diverse satellite populations. We aim to reveal some intrinsic galactic property decisive in making more or less satellites irrespective of baryonic mass or the environment in which galaxies reside.}
   {We modeled Spitzer Heritage Archive images of SAGA host galaxies at 3.6 and 4.5 microns with {\sc GALFIT} code to obtain their stellar masses. We also searched the Extragalactic Database for information on their gas content and rotation velocities.  Empirical correlations, like the baryonic Tully-Fisher relation and the stellar mass--size relation were used to exclude outliers.}
   {All the available galactic properties from the literature along with measured stellar masses were correlated with the number of satellites and no significant correlation was found. However, when we considered the "expected" number of satellites based on the correlation between the baryonic bulge-to-total ratio and the number of satellites confirmed for several nearby galaxies then strong correlations emerge between this number and (1) the mass of the bulge, and (2) the total specific angular momentum. The first correlation is positive, implying that galaxies with more massive bulges have more satellites, as already confirmed. The second correlation with the angular momentum  is negative, meaning that, the smaller the angular momentum, the greater the number of expected satellites. This would imply that either satellites cannot form if galaxy angular momentum is too high, or that satellites form inside-out, so that angular momentum is being transferred to the outer parts of the galaxies. However, deeper spectroscopic observations are needed to confirm these findings, because they rely on the expected rather than detected number of satellites. There was a luminosity limit to the SAGA survey equivalent to the luminosity of Leo I dwarf satellite of the Milky Way galaxy (the SAGA limit). In particular, correlations found in this work are very susceptible to the total number of satellites of the NGC~4158 galaxy. This galaxy is predicted to have many more satellites than detected up to the SAGA limit.}
   {}

\titlerunning{Correlations with the number of satellites}
\authorrunning{Vudragovi\'c et al.}

   \maketitle
%
\section{Introduction}

According to the standard cosmological cold-dark-matter-dominated Universe with a cosmological constant ($\Lambda$CDM model), the structures formed hierarchically, resulting in massive galaxies that had grown through accretion of or merging with smaller ones. Today, massive galaxies should be surrounded by numerous smaller dark matter halos that harbor satellites. A typical Milky Way galaxy should be able to gravitationally bind any low-mass object interior to 250-300 kpc. A direct expectation would be a positive correlation between the number of satellites and the total mass of the host galaxy, because more massive host halos  harbor more subhalos \citep{I2013}.

Although the $\Lambda$CDM model is very successful in explaining the large-scale structure distribution \citep{Eisenstein2005}, on smaller, more local scales, there are many challenges revealed in the Local Group, such as the missing satellites problem \citep{Klypin1999},  the too-big-to-fail problem \citep{BK2011}, and the plane of satellite problem \citep{Pawlowski, Muller}. Galaxies in the Local Group are close enough to enable deep observations of faint structures in detail. Another recent challenge to the $\Lambda$CDM was posed when empirical correlations were found between the number of satellites and:  (1) the mass of the bulge \citep{Javanmardi2019}, and (2) the bulge-to-total ratio \citep{Javanmardi2020} for several nearby galaxies. These two correlations are highly unlikely in the $\Lambda$CDM model, as the number of satellites should only correlate with the mass of the dark matter halo and therefore with the rotation velocity, which is closely related to dynamical mass. \citet{Javanmardi2019} used data from Milennium-II simulations \citep{BK2009} to test the significance of these empirical correlations with the predictions of the $\Lambda$CDM model, and found no significant correlation. For the former correlation, the linear correlation coefficient from simulations was measured to be 0.32, and for the latter, the correlation coefficient measured was 0.13. Empirical correlations are established for only a few nearest galaxies, and therefore a deep survey is needed to complete satellite population around Milky Way-like galaxies.
There is also the question of whether the number of satellites is correlated to some other fundamental galactic property. This information would suggest additional mechanisms guiding galaxy evolution.


Great effort has been made to complete a satellite survey of a large sample of Milky Way-like galaxies outside the Local Group. The Satellites Around Galactic Analogs (SAGA) survey\footnote{\url{http://sagasurvey.org}} \citep{Geha2017, Mao2021} is was conceived in order to discover satellite galaxies around 100 Milky Way analogs brighter than M$_{r,0} = -12.3$ (the SAGA limit, hereafter). At the present time, at Stage II of the project, 36 SAGA galaxies have been surveyed within their virial radii ($\sim$ 300 kpc) where all plausible candidate satellites have been spectroscopically confirmed. 
Host galaxies (SAGA galaxies, hereafter) are morphologically diverse, ranging from early-type (ellipticals and lenticulars) to late-type galaxies (Sb and Sc). 

The sample is large enough to enable the characterization of  a luminosity function of satellites, but also allows us to test correlations between the number of satellites and the physical properties of their host galaxies. This sample can also be used to test predictions of the  $\Lambda$CDM model on small scales outside of the Local Group. 

In Section \ref{fund_prop}, we analyze the dependence between the baryonic mass of galaxies and their near-infrared  (NIR) flux, discuss the dynamical mass estimation, and present measurements of angular momentum based on the contribution of individual galactic components (bulge and disk). In Section \ref{data}, we give a description of observational data used for the study: NIR images that were used to infer stellar mass and data retrieved from various databases on gas mass and galaxy kinematics. In Section \ref{results}, we present the correlations between the number of satellites and fundamental galactic properties of their host galaxies, and finally in Section \ref{discussion}, we discuss the implications of the correlations found on galaxy formation and evolution and present our conclusions.


\section{Fundamental galactic properties}
\label{fund_prop}

There are several fundamental galactic properties that are nontrivial to estimate: baryonic mass, dynamical mass, and total specific angular momentum. Baryonic mass, as a sum of the stellar mass and gas mass, has been shown to be more important than the stellar mass itself \citep{McGaugh2000}. Moreover, it has been shown to correlate, through the bulge-to-total baryonic mass ratio (B/T), with the total number of satellites \citep{Javanmardi2020}. The most difficult to estimate is the stellar mass, because it is measured from luminosity that cannot be converted to mass in a straightforward manner. 

Dynamical mass reflects the overall gravitational potential of galaxies, including the contributions from baryonic mass and dark matter. Assuming the rotation velocity dominates velocity dispersion in the outer parts of the galaxy, then the gravitational force balances the centrifugal force, and we can estimate an enclosed mass within the radius at which rotation velocity is measured. Single-dish HI spectra are available for a large number of nearby galaxies, and therefore the width of the HI spectral line is used as a proxy for rotation velocity. The radius of the HI disk is often several times larger than the optical radius and therefore gives a very good estimate of circular motions at large distances from the galaxy center. 

The angular momentum of galaxies is conserved during the collapse of material within a hierarchical framework of galaxy formation, but is systematically lower in early-type galaxies than in spirals \citep{Romanowsky2012}. This can be explained by mergers, whereby the angular momentum from central regions of galaxies is redistributed to their outer parts by dynamical friction. Therefore, in early-type galaxies, angular momentum could be locked up in the outer envelope populated with satellites. 

\subsection{Baryonic mass}
\label{Mbar}
Gas mass is mostly neutral hydrogen ($\mathcal{M}_{\rm HI}$) with small contributions from heavier elements. Baryonic mass is the sum of stellar mass ($\mathcal{\mathcal{M}_{*}}$) and the gas mass ($\mathcal{M}_{\rm HI}$): 
\begin{equation}
   \mathcal{M}_{\rm bar}  = \mathcal{\mathcal{M}_{*}} + 1.33 \mathcal{M}_{\rm HI}
,\end{equation}
where the contribution of heavier elements is accounted for by the factor 1.33 \citep{Courteau}.
The mass of neutral hydrogen is measured as an integral under the HI line spectrum; that is, the total flux ($F_{\rm HI}$) multiplied by the distance ($D$) squared:
\begin{equation}
    \label{eq:hi_mass}
    \frac{\mathcal{M}_{\rm HI}}{\mathcal{M}_{\odot}} = 2.33\ 10^5\ \left[\frac{F_{\rm HI}}{\rm Jy\  km\ s^{-1}}\right] \left[\frac{D}{\rm Mpc}\right]^2.
\end{equation}

Estimating the stellar mass of a galaxy is a complex and widely debated problem. There are two main approaches, which rely on extraction of the visible (baryonic) mass from the dynamical mass estimate, or on the stellar population models that correlate the observable properties, such as luminosity and color, to the galaxy stellar mass. Colors can be measured from the optical part of the spectrum or, more favorably, the NIR part, which is dominated by the light of an old stellar population that makes the bulk of the stellar mass, and is least susceptible to extinction.

In distant galaxies, where individual stars cannot be resolved, the integral light coming from both young and old stars is dominated by that from the young stars in the visible part of the spectrum, while old stars dominate the stellar mass. This problem can be solved by using the NIR light, which favors the old stellar population. Therefore, from the luminosity in the NIR (Spitzer 3.6 $\mu$m), stellar mass can be estimated as $\mathcal{M_*} = \Upsilon_{*}^{[3.6]} L_{[3.6]}$. The NIR mass-to-light ratio $\Upsilon_{*}^{[3.6]}$ needs to be constant with a typical value of $0.47 \mathcal{M_{\odot}}/L_{\odot}$ to achieve self-consistency with optical observations \citep{McGaugh2014}. The mass-to-light ratio can vary throughout the galaxy and thus affect the mass estimation of individual galaxy components, such as the  bulge and disk. \citet{Eskew2012} calibrated 3.6 and 4.5 $\mu$m fluxes to measure the stellar mass using the spatially resolved star formation history of the Large Magellanic Cloud \citep{HZ}:
\begin{equation}
    \label{eq:eskew}
    \mathcal{M_*} = 10^{5.65}\ F_{[3.6]}^{\ 2.85}\  F_{[4.5]}^{\ -1.85}\ \left[\frac{D}{0.05}\right]^2, 
    \end{equation}
where $F_{3.6}$ and $F_{4.5}$ are fluxes in 3.6 and 4.5 $\mu$m measured in Jy, $\mathcal{M_*}$ is in solar masses, and $D$ is a distance in Mpc. The scatter they obtained corresponds to 30\% uncertainty in the mass estimate. 

Their correlation is important because it provides 
a way to  estimate stellar mass for a large sample of galaxies with  NIR photometry taken by Spitzer \citep{Sheth2010} or WISE \citep{Wright2010} telescopes. In the NIR, there is only marginal sensitivity to young stellar populations, and measurements can be only slightly effected by dust. 
We therefore use the above empirical formula to measure stellar masses of SAGA galaxies from the Spitzer NIR images.

\subsection{Dynamical mass}
\label{dyn}
The dynamical mass of a galaxy with a rotationally supported HI disk is calculated as \citep{Yu2020}:
\begin{equation}
   \label{eq:dyn_mass}
   \mathcal{\frac{M_{\rm dyn}}{M_{\odot}}} = 2.31 \times 10^5 \left[\frac{V_{\rm rot}}{\rm {km\ s^{-1}}}\right]^2 \left[\frac{R_{\rm HI}}{\rm {kpc}} \right],
\end{equation}
where $R_{\rm HI}$ is the radius of HI disk. As we have no information on HI disk size from single-dish HI spectra, we use an empirical correlation \citep{Wang2016} that holds for gas masses over a large range ($10^{5.5} < M_{\rm {HI}}/M_{\odot} < 10^{10.5}$): 
\begin{equation}
    \log \left(\frac{R_{\rm HI}}{\rm {kpc}}\right) = 0.51 \log \left(\mathcal{\frac{M_{\rm HI}}{M_{\odot}}}\right) - 3.59.
    \label{eq:RHI}
\end{equation}

A better estimate of dynamical mass comes from detailed modeling of rotation curves. However, from single-dish HI spectra, only the HI line width can be measured and converted to rotation velocity applying the inclination correction.

\subsection{Specific angular momentum}
\label{am}
An expression for the specific stellar angular momentum valid for spiral galaxies was revised by \citet{Romanowsky2012} and in the case of an infinitely thin disk with an exponential radial profile is given by: 
\begin{equation}
    \label{eq:jd}
   j_d = 2 \left[\frac{V_{\rm rot}}{\rm {km\ s^{-1}}}\right] \left[\frac{R_h}{\rm {kpc}} \right],
\end{equation}
where $R_h$ is disk scale length, and $V_{\rm rot}$ is rotation velocity. SAGA galaxies have different morphologies, but exponential disks dominate their outer radial profiles. For the inner parts of galaxies, where the bulge is present, the expression from \citet{Romanowsky2012} is valid for an arbitrary value of Sersic index $n$:
\begin{equation}
    \label{eq:jb}
   j_b = k_n\ \left[\frac{v_s}{\rm{km\ s^{-1}}}\right] \ \left[\frac{R_{\rm eff}}{\rm{kpc}}\right],
\end{equation}
where $k_n\simeq 1.15+0.029\ n + 0.062\ n^2$, and $R_{\rm eff}$ is the effective radius of the bulge. The rotation velocity $v_s$  of the bulge can be calculated using the following prescription from \citet{Romanowsky2012}:

\begin{equation}
    \label{eq:vs}
  v_s = \left( \frac{v}{\sigma}\right)^*\ \sigma_0 \sqrt{\frac{\epsilon}{1-\epsilon}},
\end{equation}
where $(v/\sigma)^*$ stands for the contribution of rotational over pressure support in the galaxy dynamics. \citet{Romanowsky2012} derived a median value of $(v/\sigma)^*=0.7 \pm 0.4$ which can be applied as a single value for large sample of galaxies. Another two parameters are the  central velocity dispersion in ${\rm km\ s^{-1}}$ ($\sigma_0$) and ellipticity ($\epsilon$).

Knowing the ratio of the bulge mass to the total stellar mass, that is, the stellar B/T ratio, the total specific angular momentum \citep{Romanowsky2012} can be expressed as:
\begin{equation}
    \label{eq:j}
    j = B/T\cdot j_b + (1-B/T)\cdot j_d,
\end{equation}
where $j_b$ is the bulge angular momentum estimated from Eq.~\ref{eq:jb} and $j_d$ is the disk angular momentum given by Eq.~\ref{eq:jd}. The stellar B/T ratio in the previous equation reflects the relative contribution from a particular structural component to the total angular momentum. In the case where the stellar B/T ratio is small (B/T$<0.1$), the disk component dominates, and the bulge contribution can be neglected \citep{Romanowsky2012}.

\section{Data and sample selection}
\label{data}
The Spitzer Heritage Archive\footnote{\url{https://sha.ipac.caltech.edu/applications/Spitzer/SHA/}} contains 32 out of 36 SAGA galaxies in both 3.6 and 4.5 $\mu m$-bands. Archived images are fully reduced but not background subtracted. We estimated and subtracted an average of the median sky values inside a dozen boxes of  10 by 10 arcseconds in size far away from the galaxy. We then ran the {\sc IRAF} {\tt ellipse} task to extract radial surface brightness profiles for all the galaxies. The majority of them were part of the Spitzer Survey of Stellar Structure in Galaxies (\citealp[S4G;][]{}\citealp{Sheth2010,  MunozMateos2013, Querejeta2015}). The S4G sample includes all but three of the SAGA galaxies analyzed here (NGC~6909, NGC~7029, and NGC~5869).

In order to make a direct comparison of the results from the literature with our results, we list here the choices of selections and cuts used. Our sample is based on the SAGA survey and all SAGA galaxies were selected as isolated galaxies, meaning that there are "no galaxies brighter than $M_{\rm K} + 1$\footnote{
$M_{\rm K}$ is the total (absolute) magnitude in the NIR K-band from the Two Micron All Sky Survey (2MASS).} within 1\textdegree\ of the host, and such that the host is not within two virial radii of a massive ($5\times 10^{12} M_{\odot}$) galaxy in the 2MASS group catalog \citep{2mass}" \citep{Geha2017}.  
As \citet{Javanmardi2019} and \citet{Javanmardi2020} compared nearby galaxies including the most massive Local Group galaxies to the Milennium-II simulation results, their selection criteria favor halos with virial masses of between $10^{10}$ and $2 \times 10^{12}$ M$_{\odot}$ that are isolated and harbor central disk galaxies. The sample of nearby galaxies to which the simulations are compared in this work comprises: the Milky Way, M101, M31, M33, M94, M81, and CenA. These span the following range of stellar mass $1 \times 10^{10} - 11.2 \times 10^{10}$ M$_{\odot}$, and a full range in bulge-to-total ratio ($0.01 < B/T < 1$). The stellar mass of SAGA galaxies goes from $1.8 \times 10^{10}$ to $11.8 \times 10^{10}$ M$_{\odot}$, and bulge-to-total ratios cover values from 0.01 to 0.8. The only caveat to keep in mind is the brightness limit of the satellite populations. Nearby galaxies from the studies of \citet{Javanmardi2019} and \citet{Javanmardi2020} have the satellites brightness limit of $M_{V} \leq -8.2$ mag, while the satellites of the SAGA galaxies  are all brighter than M$_{r,0} = -12.3$ mag, which is the SAGA limit. In the remainder of the paper, we remove all the satellites fainter than the SAGA limit from the sample of nearby galaxies \citep{Javanmardi2019, Javanmardi2020}  for direct comparison, and use an empirical correlation found between the number of satellites and the bulge-to-total ratio to recover the faint end of the satellite population for SAGA galaxies to enable direct comparison with the total number of satellites.

First, we inspected the inclinations of all SAGA galaxies and removed the three found to be edge-on (NGC~4348, PGC~13646, and UGC~903). The remaining 29 galaxies were used to infer the relation between bulge-to-total ratio and the number of satellites. Hereafter in our analysis, as it is strongly affected by the accuracy of the distance and/or rotation velocity measurements, we applied two empirical relations (stellar mass--size and the baryonic Tully Fisher relation) to remove seven outliers (NGC~1309, NGC~2962, NGC~4454, NGC~5347, NGC~5602, NGC~5869 and PGC~68743). Another two galaxies were discarded from the sample because we were not able to find the data on either their HI mass or rotation velocity, which was needed to estimate dynamical mass. For the remaining 20 galaxies dynamical mass was estimated and correlated with the number of satellites. Finally, for specific angular momentum measurement velocity dispersion was needed to estimate the contribution of the bulge to the total angular momentum, and for two galaxies we lack these data (NGC~5792 and NGC~7328). These two were therefore removed from the correlation between the number of satellites and the specific total angular momentum. For this last correlation, the total number of SAGA galaxies used is 18. These 18 galaxies have passed all empirical tests and have all the data needed. More details on these selection cuts are given in the following section, and the number of galaxies used for each of the correlations is given in  Table~\ref{num_sat}. 

\begin{table}[ht]
    \caption{\label{num_sat}Correlations examined in this work}
    \centering
    \begin{tabular}{|c|c|}
    \hline
        Correlation & Number of galaxies \\
        \hline
        B/T vs. N$_{\rm sat}$ & 29  \\ 
        $\mathcal{M}_{\rm bulge}$ vs. N$_{\rm sat}$ & 29 \\ 
        $\mathcal{M}_{\rm dyn}$ vs. N$_{\rm sat}$ & 20  \\ 
        $j$ vs. N$_{\rm sat}$ & 18  \\ \hline
    \end{tabular}
    \tablefoot{In the first column the name of the correlation is listed, while in the second column the total number of SAGA galaxies is indicated and is related to the particular correlation that has been derived. B/T is the baryonic bulge-to-total ratio, $\mathcal{M}_{\rm bulge}$ is the bulge mass, $\mathcal{M}_{\rm dyn}$ is dynamical mass, and $j$ is the total specific angular momentum. All these properties are correlated with the number of satellites N$_{\rm sat}$.}
\end{table}

\begin{figure}
\centering
\includegraphics[width = .5\textwidth]{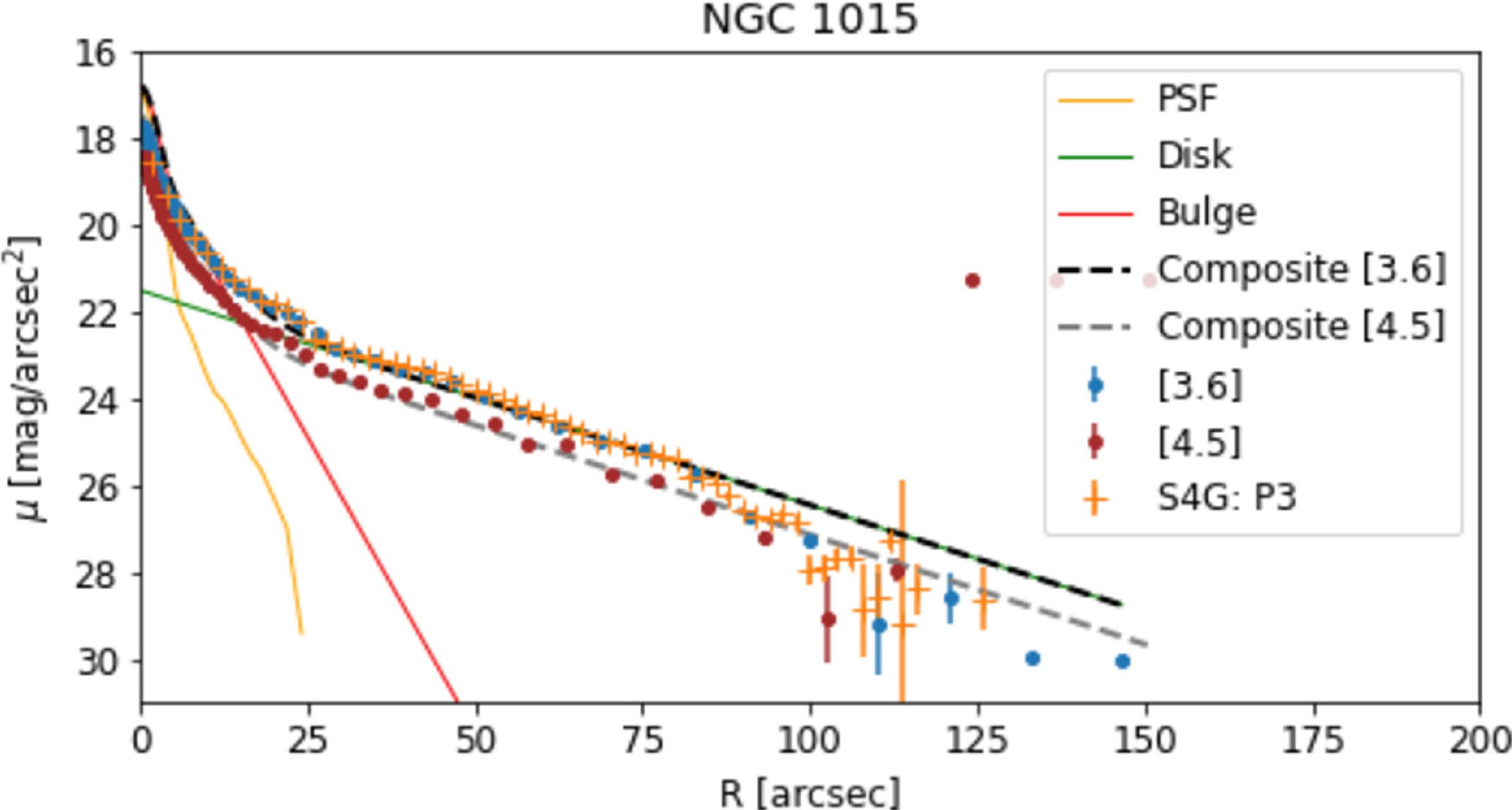}
\caption{Radial profile of the NGC~1015 galaxy in 3.6 and 4.5 $\mu$m-bands modeled with the IRAF {\tt ellipse} task, denoted as blue and red points, respectively. Overplotted are orange crosses from \citet{Munoz2015} for the 3.6 $\mu$m-band. For clarity, only composite models are given for 3.6 $\mu$m (black) and 4.5 $\mu$m bands (gray), that correspond to the sum of all individual subcomponents.  Individual subcomponents are presented only for the 3.6 $\mu$m band (point spread function, bulge and disk). 
}
\label{fig:radprof}
\end{figure}

Background-subtracted images are modeled with the {\sc GALFIT} code \citep{Peng2010} using multiple components including the disk and bulge with the point spread function (PSF) often needed to describe the nucleus, and additional components like bars and rings where needed. {\sc GALFIT} has become a standard tool for modeling galactic structure. In Fig.~\ref{fig:radprofs}.1 (Appendix~\ref{app_plots}), we plot all the individual modeled subcomponents, along with the composite models. In addition to the radial profiles in 3.6 and 4.5 $\mu$m-bands, Figure \ref{fig:radprof} presents the sum of all individual subcomponents modeled with {\sc GALFIT} for each galaxy, and is referred to here as a composite model. Individual components are the nucleus described with the point spread function (PSF), and the bulge and disk. Comparison with radial profile of S4G:P3 is indicated for the 3.6 $\mu$m band only (4.5 $\mu$m omitted) for clarity. Our radial profiles in the 3.6 $\mu m$ band show excellent agreement with the radial profiles of S4G galaxies in common with the S4G survey extracted from their Pipeline 3 \citep{Munoz2015}, as can be seen for all SAGA galaxies in Fig.~\ref{fig:radprofs}.1 (Appendix~\ref{app_plots}).

The Spitzer images are in units of M Jy sr$^{-1}$ given in the AB system. To convert total (integral) magnitudes measured by the {\sc GALFIT} code into fluxes in Jy that are required for stellar mass estimation (Eq.~\ref{eq:eskew}), we used the following conversion \citep{Sheth2010}:
\begin{equation}
    F = 10^{-0.4\ (m_{\rm AB}-8.9)},
\end{equation}
where $F$ is the flux in Jy and $m_{\rm AB}$ is the magnitude in the AB system. With this conversion,  Eq.~\ref{eq:eskew} now reads:
\begin{equation}
    \label{eq:stellar_mass}
    \frac{\mathcal{M_*}}{\mathcal{M}_{\odot}} = 64.78 \times 10^{10}\ \left[10^{-0.4\  \left(m_{[3.6]}-m_{[4.5]}\right)}\right]^{1.85}  10^{-0.4\ m_{[3.6]}}\ \left[\frac{D}{\rm{Mpc}}\right]^2.
\end{equation}
Here, distance $D$ to the source should be provided in Mpc, and total (integral) magnitudes $m_{[3.6]}$ and $m_{[4.5]}$ are in the AB system. The calculated stellar mass $\mathcal{M_*}$ is in solar masses. 

As one of the correlations that has been found for nearby galaxies by \citet{Javanmardi2020}, namely that between baryonic B/T ratio and the number of satellites, involves gas mass in addition to stellar mass (i.e., baryonic), we searched the Extragalactic Distance  Database (EDD)\footnote{https://edd.ifa.hawaii.edu} \citep{EDD} to complement our stellar mass measurements with gas mass estimates. Most of the galaxies are cataloged in the EDD catalog "All Digital HI" with HI flux measured. The rest of the galaxies are listed in the EDD catalog "Pre-Digital HI"\footnote{"Pre-Digital HI" catalog is a catalog of pre-digital HI line-width parameters collected by Tully.}, where the measurements of $\log F_{\rm {HI}}$ flux are given. Three galaxies are also part of the ATLAS3D project \citep{atlas}: NGC~2962, NGC~5869, and NGC~6278. For these three galaxies, we retrieved data on H$_2$ mass.  
Combining stellar mass with gas mass (atomic and molecular)  provides us with the total baryonic mass. 

From the EDD catalog "All Digital HI" we retrieved the averaged width of the HI line (Wmx$_\mathrm{av}$). The width of the HI line can be corrected to give maximum rotation velocity, which can be used to infer the specific stellar angular momentum and dynamical mass of galaxies. To convert Wmx$_\mathrm{av}$ into rotation velocity $V_\mathrm{rot}$=Wmx$_{\mathrm{av}}/2\sin(i)$, we need to correct for inclination, which has been estimated from the ratio of the minor to the major galaxy disk axis (b/a): $\sin^2(i) = 1 - [{(b/a)}^2 - q_0^2]/[1-q_0^2]$, where $q_0=0.11$ \citep{G1992}. For three galaxies that were part of the ATLAS3D project \citep{atlas}, namely NGC~2962, NGC~5869, and NGC~6278, rotation velocities were taken from \citet{Krajnovic2011}.

\section{Results}
\label{results}

We modeled all SAGA galaxies in the 3.6 $\mu$m band with multiple components using the {\sc GALFIT} code. The results are presented in Appendix~\ref{app_table} and \ref{app_plots} (table and surface brightness profiles, respectfully). The baryonic mass is measured as a sum of stellar and gas mass: $\mathcal{M}_{\rm} = \mathcal{M_{*}}+1.33\ \mathcal{M}_{\rm HI}$. After checking galaxy inclinations, which were estimated from the ratio of the disk minor to major axis from {\sc GALFIT} disk models, we excluded three galaxies as edge-on, because this inclination is unfavorable for bulge mass estimation (NGC~4348, PGC~13646, and UGC~903).


\begin{figure}
\centering
\includegraphics[width = .5\textwidth]{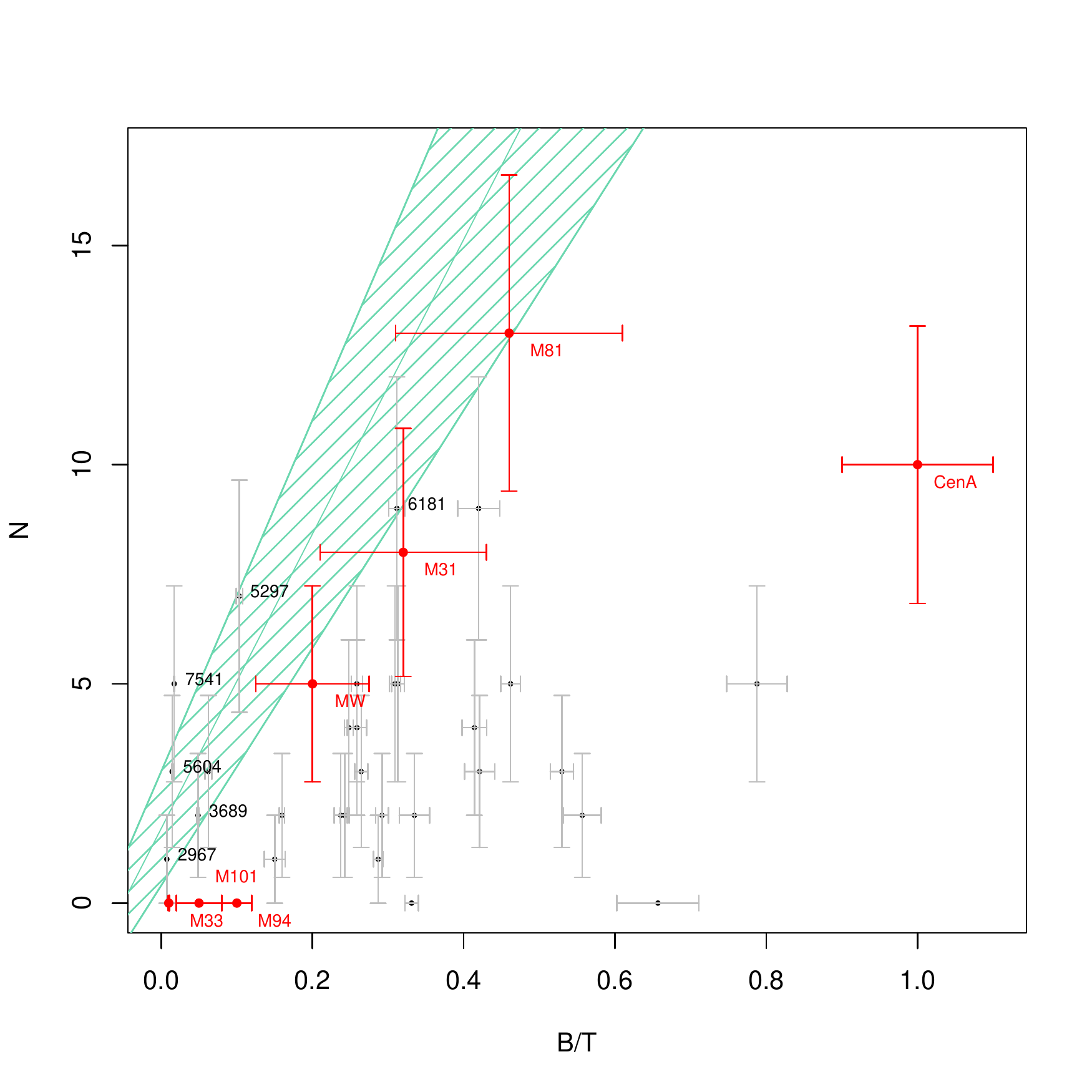}
\caption{Relation between the B/T ratio and the number of satellites brighter than $M_{r,0}<-12.3$ (SAGA limit). Red points are taken from \citet{Javanmardi2020} and references therein. Their values on the y-axis are corrected to represent the number of satellites brighter than the SAGA limit. Gray points are measured in this work. The green shaded area is the correlation between bulge-to-total ratio and the number of satellites from \citet{Javanmardi2020}, which is found for the total number of satellites for galaxies MW, M101, M33, M31, CenA, M81, and M94.}
\label{fig:bt_relation_new}
\end{figure}

\begin{figure}
    \centering
    \includegraphics[width = .5\textwidth]{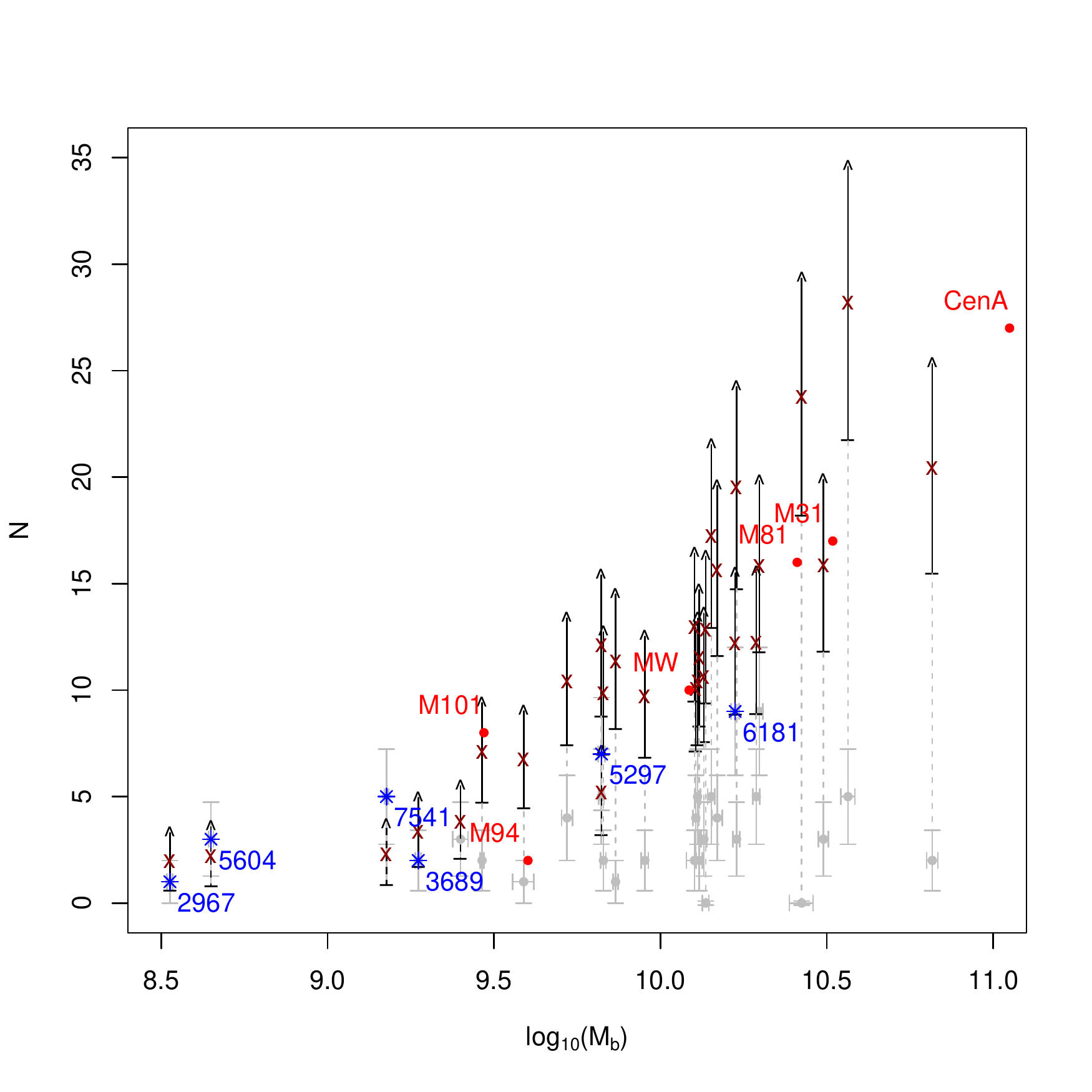}
    \caption{Bulge mass vs. expected number of satellites for SAGA galaxies (dark red crosses and blue asterisks) and the number of satellites brighter than the SAGA limit (gray circles). Overplotted are the galaxies from \citet{Javanmardi2020} with their total number of satellites as red circles. Blue asterisks indicate SAGA galaxies that according to the relation between the B/T ratio and the number of satellites found in \citet{Javanmardi2020} have a complete (or nearly) complete census of satellites.}
    \label{fig:mb_nsat}
\end{figure}

For each galaxy component (bulge and disk), total mass  depends on the distance squared (Eqs.~\ref{eq:hi_mass} and \ref{eq:stellar_mass}), while B/T ratio as the ratio of masses is not affected by distance uncertainties. 
To enable comparison with \citet{Javanmardi2020}, we excluded all the satellites fainter than $M_{\rm r,0}<-12.3$ for all the host galaxies from \citet{Javanmardi2020} (and references therein). In the paper by \citet{McConnachie2012}, we count five satellites of the Milky Way and eight satellites of the Andromeda galaxy brighter than the SAGA limit and up to 250 kpc. For the M81 galaxy, we refer to \citet{Chiboucas2013}, where galaxies from the M81 group are listed with their physical separation and absolute magnitudes. Within 250 kpc from M81 galaxy, we count 20 satellites brighter than the SAGA limit. \citet{Crnojevic2019} list all satellites of CenA galaxy within 250 kpc, and we count 10 satellites brighter than the SAGA limit. M94 galaxy has been observed by \citet{Smercina2018} but only up to 150 kpc distance, and only two satellites had been detected down to M$_{\rm V}$ = -9.1 mag. As they are both fainter than our limit, we  consider that M94 has no satellites. M33 galaxy has one possible but unconfirmed satellite fainter than M$_{\rm V}$ = -6.5 mag, as reported by \citet{Martin2009}, and therefore we consider it has no satellites. M101 galaxy has eight confirmed satellites in total \citep{Carlsten2019, Bennet2019}, but they are all fainter than the SAGA limit.

In Fig.~\ref{fig:bt_relation_new}, SAGA galaxies are given in gray along with their error bars (Table \ref{tab:bt_table}), while galaxies from \citet{Javanmardi2020} are given in red. As discussed above, for galaxies from \citet{Javanmardi2020} (red circles in the Fig.~\ref{fig:bt_relation_new}) the number of satellites is limited to those brighter than the SAGA limit to enable direct comparison. If we assume that the relation holds, then all the galaxies that lie below the green area have numerous satellites that were missed by the SAGA survey. We cannot confirm the validity of the relation with these data, because we only include satellites brighter than a certain limit. However, knowing that, when the total number of satellites is taken into account, all red points from \citet{Javanmardi2020} enter the green shaded area, there is a possibility that the relation truly holds. Another sign that the relation may hold is that there are no points above the green shaded area, the presence of which would question the correlation. Simply, there can be only more satellites than detected, not less. All the points can only go up, never down, if the correlation holds, and this is satisfied with the existing data. We label six SAGA hosts that enter the green shaded area with their NGC numbers, and we consider they have already a total number of satellites detected, and thus are directly comparable to the sample of nearby galaxies with complete satellite census. For these six SAGA galaxies, and those from \citet{Javanmardi2020} with total number of satellites, we measure a Pearson's linear correlation coefficient of 0.73.

Another correlation that is linked to the relation between B/T ratio and the number of satellites is the correlation between the mass of the bulge and the number of satellites \citep{Javanmardi2019}. For each SAGA galaxy, we only have information on the number of satellites brighter than the SAGA limit and, on the other hand, \citet{Javanmardi2019} refers to the total number of satellites, and so we estimated the total number of satellites for SAGA galaxies in order to compare the results. The total number of satellites, if the relation between the number of satellites and B/T ratio \citep{Javanmardi2019} holds, is: N = 33.6 ($\pm$6.5) B/T + 1.7 ($\pm$1.3), as given in \citet{Javanmardi2020}. This correlation, along with its uncertainties, is represented as the green shaded area in Fig.~\ref{fig:bt_relation_new}. We calculated the average number of satellites for SAGA galaxies according to this relation, and refer to it as the expected number of satellites. This expected number of satellites is marked with dark red crosses with black arrows for the error bars in Fig.~\ref{fig:mb_nsat}. Overplotted are gray circles with errors referring to the number of satellites brighter than the SAGA limit ($M_{\rm r,0}<-12.3$). However, blue asterisks mark six SAGA galaxies which, according to Fig.~\ref{fig:bt_relation_new}, have a complete census of satellites (they fall into the green shaded area), and for which a direct comparison can be made with the study of \citet{Javanmardi2019} and \citet{Javanmardi2020}. Unlike Fig.~\ref{fig:bt_relation_new}, where fainter satellites were excluded, the same galaxies (red points) in Fig.~\ref{fig:mb_nsat} harbor the total number of satellites. There is a positive trend indicating that galaxies with larger (more massive) bulges tend to harbor more satellites. Galaxies that have a complete census of satellites and thus are comparable (Fig.~\ref{fig:mb_nsat}), labeled as blue asterisks from SAGA sample and red circles from \citet{Javanmardi2019}, follow this positive correlation between the mass of the bulge and the (total) number of satellites. For these galaxies with a complete satellite census (13 in total), the Pearson's linear correlation coefficient is 0.88.

\begin{table}[ht]
    \caption{\label{tab:bt_table}Baryonic bulge-to-total ratio and the number of satellites for sample galaxies.}
    \centering
    \begin{tabular}{lrrrr}
      \hline \hline
    Name & B/T & e\_B/T & N & e\_N \\ 
    \hline \hline
  \multicolumn{5}{c}{SAGA} \\
  \hline
  1015 & 0.237 & 0.009 &    2 & 1.414 \\ 
  1309 & 0.062 & 0.004 &    3 & 1.732 \\ 
  2543 & 0.292 & 0.008 &    2 & 1.414 \\ 
  2962 & 0.265 & 0.009 &    3 & 1.732 \\ 
  2967 & 0.007 & 0.001 &    1 & 1.000 \\ 
  3689 & 0.049 & 0.001 &    2 & 1.414 \\ 
  3976 & 0.557 & 0.025 &    2 & 1.414 \\ 
  4158 & 0.788 & 0.040 &    5 & 2.236 \\ 
  4454 & 0.259 & 0.013 &    4 & 2.000 \\ 
  5297 & 0.103 & 0.004 &    7 & 2.646 \\ 
  5347 & 0.160 & 0.004 &    2 & 1.414 \\ 
  5448 & 0.259 & 0.007 &    5 & 2.236 \\ 
  5602 & 0.309 & 0.007 &    5 & 2.236 \\ 
  5604 & 0.014 & 0.001 &    3 & 1.732 \\ 
  5633 & 0.331 & 0.009 &    0 & 0.000 \\ 
  5690 & 0.530 & 0.015 &    3 & 1.732 \\ 
  5750 & 0.287 & 0.006 &    1 & 1.000 \\ 
  5792 & 0.421 & 0.020 &    3 & 1.732 \\ 
  5869 & 0.657 & 0.054 &    0 & 0.000 \\ 
  5962 & 0.335 & 0.020 &    2 & 1.414 \\ 
  6181 & 0.312 & 0.011 &    9 & 3.000 \\ 
  6278 & 0.420 & 0.028 &    9 & 3.000 \\ 
  6909 & 0.414 & 0.016 &    4 & 2.000 \\ 
  7029 & 0.313 & 0.009 &    5 & 2.236 \\ 
  7079 & 0.248 & 0.006 &    4 & 2.000 \\ 
  7328 & 0.150 & 0.014 &    1 & 1.000 \\ 
  7541 & 0.017 & 0.000 &    5 & 2.236 \\ 
  7716 & 0.242 & 0.006 &    2 & 1.414 \\ 
  68743 & 0.462 & 0.013 &    5 & 2.236 \\ 
  \hline
  \multicolumn{5}{c}{\citet{Javanmardi2020}} \\
  \hline
  MW & 0.200 & 0.075 &    5 & 2.236 \\ 
  M31 & 0.320 & 0.110 &    8 & 2.828 \\ 
  M33 & 0.010 & 0.001 &    0 & 0.000 \\ 
  M81 & 0.460 & 0.150 &   13 & 3.606 \\ 
  CenA & 1.000 & 0.100 &   10 & 3.162 \\ 
  M94 & 0.100 & 0.020 &    0 & 0.000 \\ 
  M101 & 0.050 & 0.030 &    0 & 0.000 \\ 
  \hline
\end{tabular}
\tablefoot{Baryonic bulge-to-total mass ratio (B/T) and the number of satellites (N) brighter than the SAGA limit are given along with their corresponding errors for: all SAGA galaxies (upper part) labeled with their NGC number with one exception of a single five-digit number that has a PGC prefix (the last SAGA galaxy listed), and nearby galaxies from \citet{Javanmardi2020} (lower part). The number of satellites from \citet{Javanmardi2020} has been reduced to the SAGA limit for direct comparison.}
\end{table}

\begin{figure}
    \centering
    \includegraphics[width = .5\textwidth]{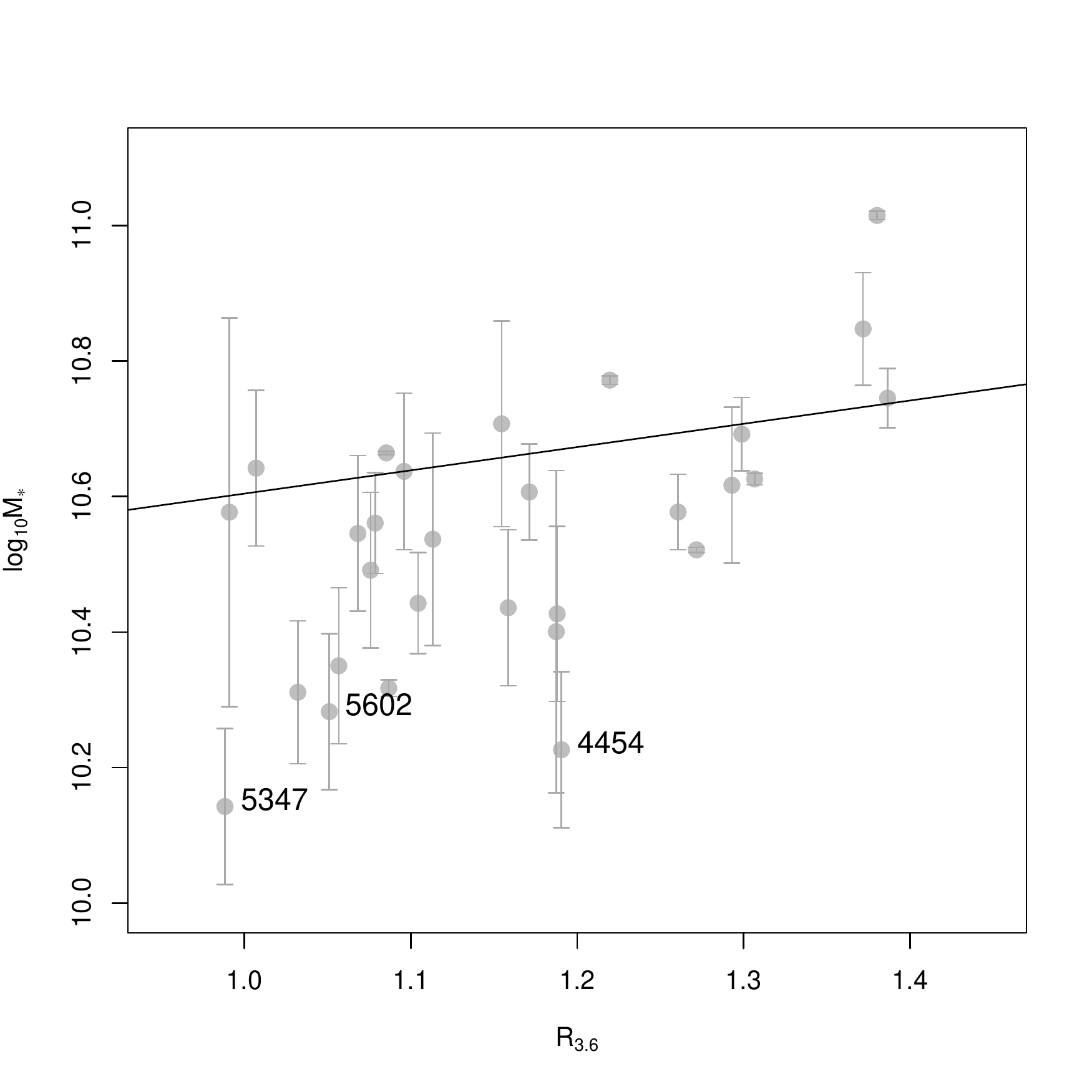}
    \caption{Stellar mass--size relation for SAGA galaxies: gray circles denote stellar mass with their corresponding errors. Stellar mass is correlated with the radius at which surface brightness at 3.6 $\mu$m drops to 25.5 AB mag/arcsec$^{-2}$ ($R_{3.6}$). Galaxies labeled with their NGC numbers are 2-sigma outliers from the fitted relation (black line).}
    \label{fig:msr}
\end{figure}

\subsection{Constraints from the empirical relations }
\label{emp}
For the other two relations we are investigating, namely number of satellites--specific angular momentum and  number of satellites--galactic dynamical mass, we need to be certain of the quality of our measurements as well as that of the measurements we acquire from various resources. 

\begin{figure}
\centering
\includegraphics[width = .5\textwidth]{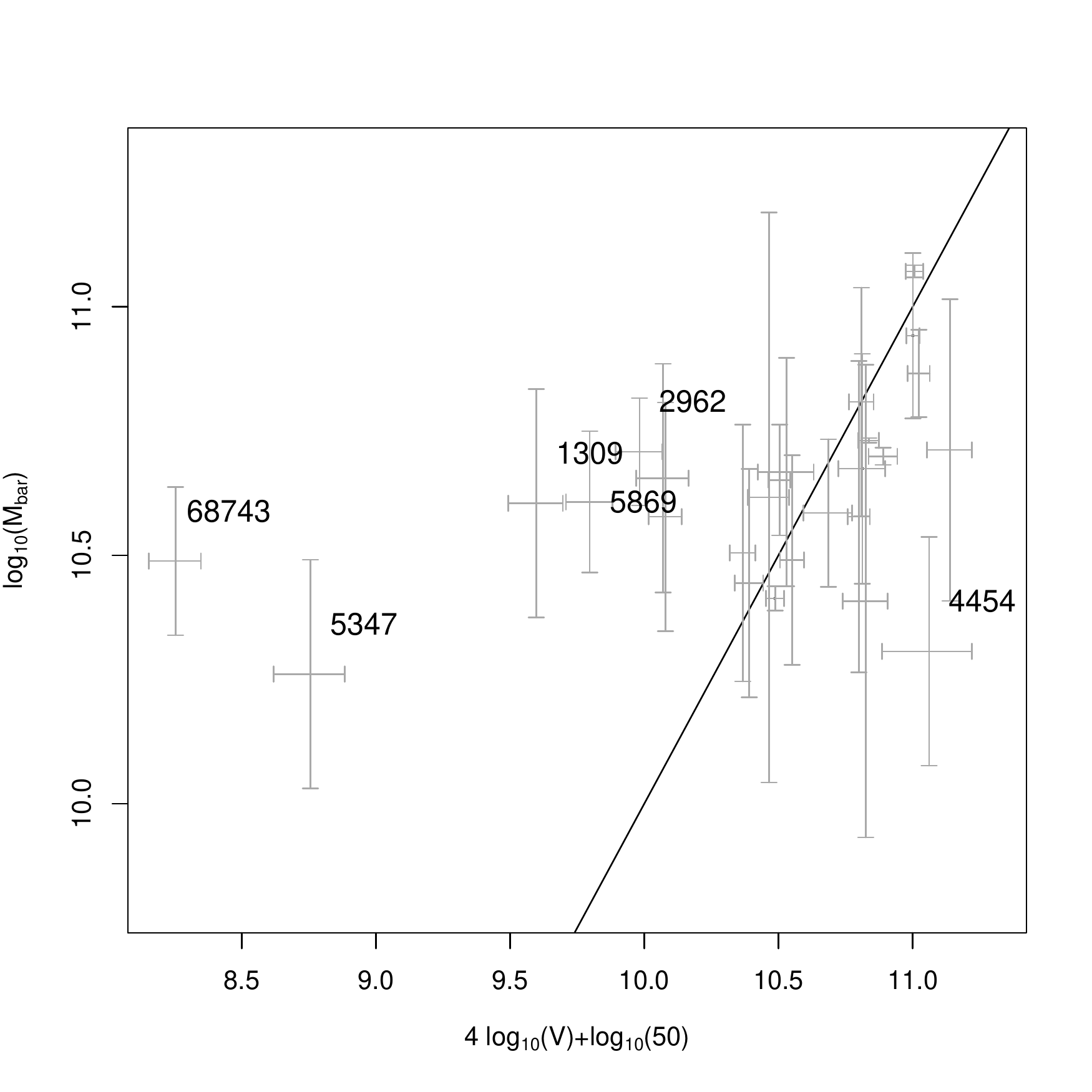}
\caption{Baryonic Tully-Fisher relation from \citet{McGaugh2005}: the solid line is simply Eq.~5 from \citet{McGaugh2005}, not the fit; 1-sigma outliers are labeled with their NGC names, omitting the NGC prefix, with the exception of five-digit number 68743, which corresponds to the PGC prefix. Other points are not labeled for clarity.}
\label{fig:btfr_all_relation}
\end{figure}

First, we checked our stellar mass measurements using the empirical relation between the radius at which the surface brightness profile of the galaxy at 3.6 $\mu$m drops to 25.5 AB mag/arcsec$^{-2}$ and its stellar mass \citep{Munoz2015}, the so-called stellar mass--size relation. For our sample of galaxies, the relation is given in Fig.~\ref{fig:msr}. Stellar mass is plotted in gray circles with corresponding error bars, while the fit is given as a black solid line. The intention was not to make another comparison, but simply to find outliers. There are three 2-sigma outliers and we shall exclude them on the basis of inaccurate mass measurements. For NGC~5347 and NGC~5602, the composite fit (Fig.~\ref{fig:radprofs}.1) departs from the radial profile at the 25.5 AB mag/arcsec$^{-2}$ level. This could be why these two galaxies fall off the relation. We cannot identify why galaxy NGC~4454 deviates from the relation, because the fit follows the radial profile. However, this galaxy along with NGC~5347 falls off the baryonic Tully-Fisher relation, as can be seen in Fig.~\ref{fig:btfr_all_relation}. This is an additional reason to exclude these two galaxies from the sample. 

Additional parameters of huge importance to our further analysis are the rotation velocity and galaxy distance. Both parameters are equally important in deriving the dynamical mass and total specific angular momentum. The rotation velocity is measured from the width of the HI spectral line (Wmx$_{\rm ave}$ taken from EDD database), which is corrected for inclination as explained in detail in the previous section. The radius of the HI disk is estimated from $\mathcal{M_{\rm HI}}$, which depends on the distance squared (Eq.~\ref{eq:hi_mass}). Another galactic property of interest to us is the total specific angular momentum, for which again the rotation velocity and an accurate distance measurement are needed. We find no information on the HI line width (and/or rotation velocity) for three of the  galaxies of our sample: NGC~7079, NGC~6909 and NGC~7029. To check both rotation velocity and distance accuracy, we employ the baryonic Tully-Fisher relation \citep{McGaugh2000, McGaugh2005}. The baryonic Tully-Fisher relation correlates the total baryonic mass and the flat part of the rotation curve, which is often very close to the maximum rotation velocity that we have estimated using the width of HI line from the EDD database. This is the linear relation that reads: $\mathcal{M/M_{\odot}}=50 \cdot V_{\rm rot}^4$. Inserting this relation into Fig.~\ref{fig:btfr_all_relation} we calculated 1-sigma deviations and labeled the galaxies that deviate more than 1-sigma with their NGC names omitting the NGC prefix, and for the single case with five digits 68743 omitting the PGC prefix (PGC~68743). All these galaxies have either questionable distance measurements or inclination corrections. In either case, we exclude them from the sample for the subsequent analysis. Two of them are already 1-sigma outliers from the stellar mass--size relation (NGC~5347 and NGC~4454). In summary, we exclude seven more galaxies (apart from three edge-on galaxies) as outliers either from the stellar mass--size relation or the baryonic Tully-Fisher relation. To independently test our rotation velocity measurements, we searched the Hyperleda database through the EDD website for all SAGA galaxies and retrieved rotation velocity. Comparing our rotation velocities with values from the HyperLeda database\footnote{\url{http://leda.univ-lyon1.fr}}, we find three 1-sigma outliers: NGC~1309, NGC~2967, and NGC~4158. This is in line with our previous findings, except for the galaxy NGC~2967, which we decided to keep because it is one of the six galaxies that according to \citet{Javanmardi2020} have a complete census of satellites.

\subsection{Correlations with fundamental galactic properties}

\begin{figure}
    \centering
    \includegraphics[width = .5\textwidth]{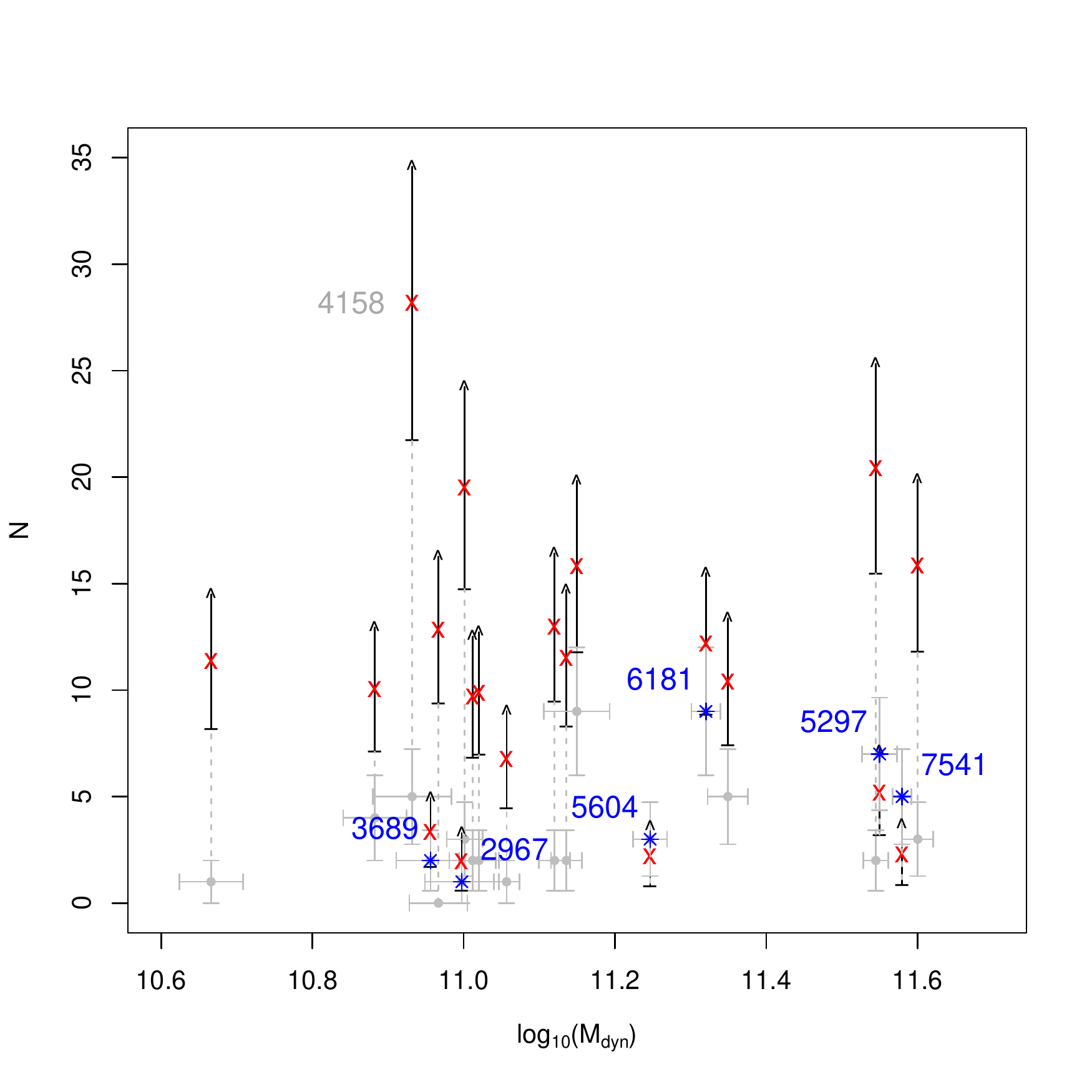}
    \caption{Relation between dynamical mass and the number of satellites for SAGA galaxies, for which HI line measurements exist in the EDD database: gray circles with error bars denote the number of satellites detected up to the limit of $M_{\rm r,0}<-12.3$; red crosses give the expected number of satellites if the relation from \citet{Javanmardi2019} holds. Error bars on the y-axis are simply the square root of the number of satellites. Dynamical mass is calculated from Eq.~\ref{eq:dyn_mass}, with error bars corresponding to the errors on the rotation velocity only, meaning they are underestimated.}
    \label{fig:dyn_mass}
\end{figure}

For dynamical mass measurements (Eq.~\ref{eq:dyn_mass}), both the rotation velocity and the radius of the HI disk are needed. For the galaxy NGC~5869, there are no measurements on its HI mass, and so we were not able to use Eq.~\ref{eq:dyn_mass} to infer its dynamical mass from the neutral gas. This galaxy is omitted from Fig.~\ref{fig:dyn_mass}, where we plot dynamical mass versus the average number of satellites expected from the relation between the B/T ratio and the number of satellites found in \cite{Javanmardi2020}. The dark red crosses in Fig. 6 show the expected number of satellites given their B/T ratios if this latter relation holds for our sample of galaxies. Uncertainties are denoted with arrows.
The number of bright satellites detected in the SAGA survey are presented as gray crosses which stand for error bars. The blue asterisks labeled with their NGC names are the only six galaxies in our sample that are considered to have a complete satellite census. Our analysis indicates that these six galaxies should stay in place, while all other galaxies (gray points) should migrate upwards toward red crosses, implying they have more satellites than detected. 
The inspected range of dynamical mass is rather narrow, but it shows no correlation, unless it is buried in the large scatter.

From each bulge component, we extracted information on the effective radius, Sersic index, and the ratio of minor to major axis (as $\epsilon=1-b/a$) from the best fit model (Table \ref{tab:all_data_app}) to measure its specific angular momentum (Table ~\ref{tab:alldata}) using Eqs.~\ref{eq:jb} - \ref{eq:vs}. Velocity dispersion was mainly taken from the SDSS SkyServer CasJobs Service\footnote{\url{https://skyserver.sdss.org/casjobs/}} where galaxies can be cross-matched against the SpecObjAll catalog with measured velocity dispersion based on their celestial coordinates. We find 18 matches within 0.5 arcmin radius from the galaxy center. However, the fiber is placed on the galaxy center for only ten of these matches. Three galaxies (NGC~2962, NGC~5869, and NGC~6278) were found in the catalog of  \citet{Dabringhausen2016}. We searched the HyperLeda database for the rest of the sample. As a result, data on the central velocity dispersion were gathered for all but six galaxies: NGC~7328, NGC~3689, NGC~5604, NGC~5792, NGC~4454, and NGC~5602. NGC~4454 and NGC~5602 are already excluded as outliers from empirical relations. Of the four remaining galaxies missing information on the central velocity dispersion, two of them (NGC~3689 and NGC~5604) have a stellar B/T ratio of less than 0.1, which makes the bulge angular momentum contribution negligible compared to the disk component, and these two galaxies are kept for further study. Galaxies NGC~7328 and NGC~5792 are excluded because the contribution of the bulge component to the total angular momentum cannot be neglected (stellar B/T $\ge$ 0.2), and thus the specific angular momentum of the disk cannot well represent the total angular momentum, contrary to the previous two cases of NGC~3689 and NGC~5604 galaxies. All the structural data measured from Spitzer's images and those additional data retrieved from databases used for angular momentum estimation are listed in Table \ref{tab:all_data_app}.1 (Appendix~\ref{app_table}).

\begin{figure}
    \centering
    \includegraphics[width = .5\textwidth]{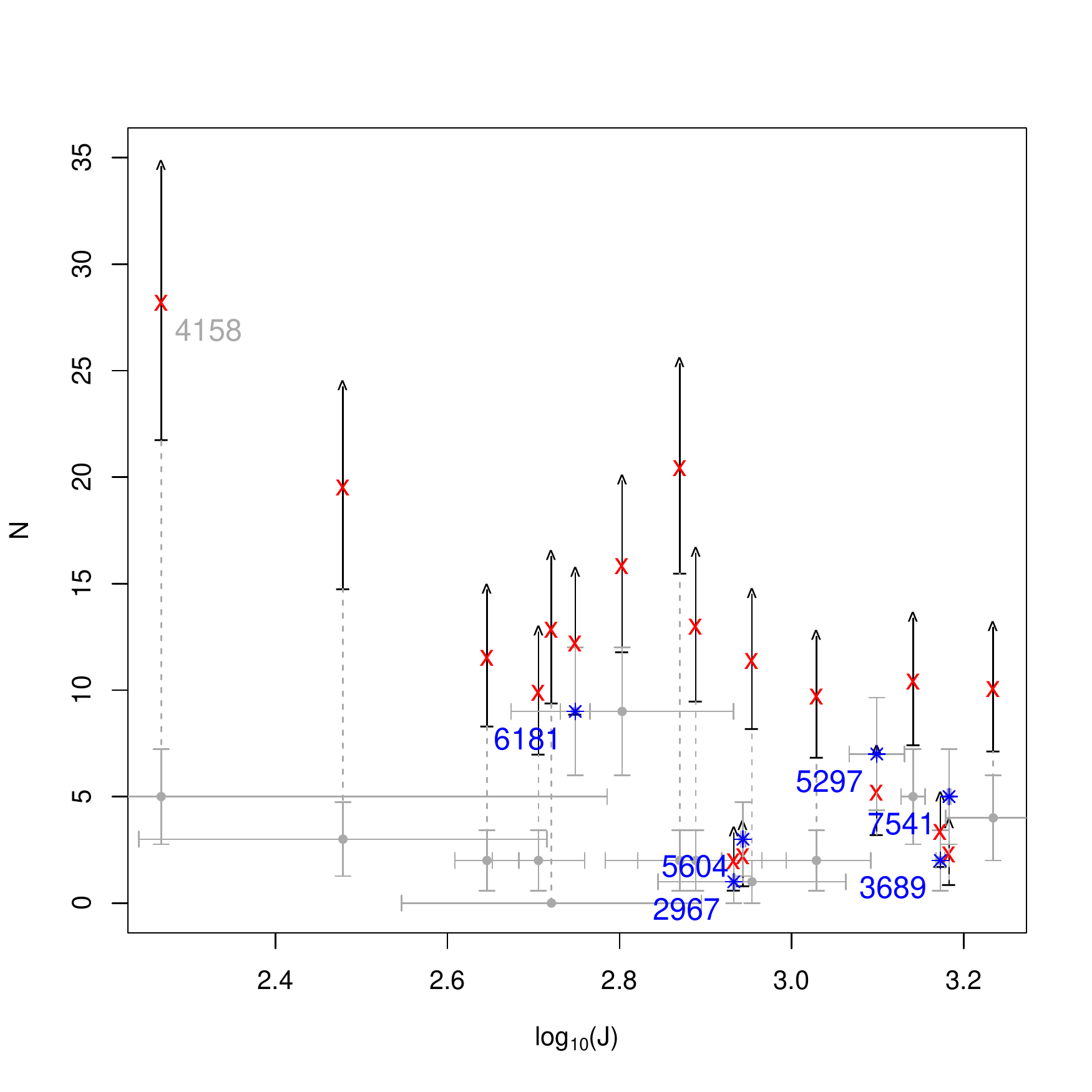}
    \caption{Relation between the total specific angular momentum (Eq.~\ref{eq:j}) and the number of satellites for SAGA galaxies, for which HI line measurements exist in the EDD database: gray circles with error bars denote the number of satellites detected up to the SAGA limit ($M_{\rm r,0}<-12.3$); red crosses give the expected number of satellites if the relation from \citet{Javanmardi2019} holds.}
    \label{fig:j}
\end{figure}

From {\sc GALFIT} exponential disk models, we extracted disk scale lengths and converted them into physical units (kpc) using galaxy distances from the SAGA dataset \footnote{\url{http://sagasurvey.org}} to measure the specific stellar angular momentum of the disks (Eq.~\ref{eq:jd}). With additional data on maximum rotation velocity and stellar B/T ratio, we measured specific angular momentum for 18 galaxies in total. The correlation between total specific angular momentum and the expected number of satellites is shown in Fig.~\ref{fig:j}. The expected number of satellites for each galaxy is marked with a dark red cross in Fig.~\ref{fig:j} (Table~\ref{tab:alldata}), with the vertical arrow representing the error bar. The number of satellites detected in the SAGA survey is denoted with a gray circle with error bars holding the errors on both axes.  The Pearson coefficient for the correlation is significant: -0.74. A single outlier in the Fig. \ref{fig:dyn_mass} is the galaxy NGC~4158, which is also the main driver of the correlation with angular momentum (Fig.~\ref{fig:j}). Thus, deeper spectroscopic observations of fainter satellites would help us to decide whether this correlation truly holds. This galaxy should have 23 satellites fainter than the SAGA limit. 




\section{Discussion and conclusions}
\label{discussion}
All galaxies studied here are part of the SAGA survey as Milky Way analogs. We searched for previous and new correlations between the number of satellites and fundamental galactic properties to understand  why some galaxies that are very similar regarding their stellar mass and environment have quite different numbers of satellites. Environment plays no role in their satellite distribution function because they are all selected as isolated galaxies.

The $\Lambda$CDM model predicts an excess of bright satellites compared to observations (the "too-big-to-fail" problem; \citet{BK2011}), and predicts no significant correlations between the number of satellites and the mass of the bulge or the bulge-to-total ratio \citep{Javanmardi2019, Javanmardi2020} for the Milky Way-like galaxies. However, discrepancies were mainly obtained for Local Group galaxies. The necessity for a SAGA sample of galaxies comes from the concern that Local Group satellites are representative of typical galaxies in this mass range. The correlation between the bulge-to-total ratio \citep{Javanmardi2020} and the number of satellites holds for the galaxies with complete satellite census, which helped us identify six such galaxies in the SAGA sample. We confirmed a very strong correlation between the mass of the bulge and the number of satellites \citep{Javanmardi2019} for these galaxies, including the sample of nearby galaxies for which the total number of satellites is known. We used the former correlation to predict the total number of satellites for SAGA galaxies based on their bulge-to-total ratio. This expected number of satellites makes the correlation between the mass of the bulge and the number of satellites even more significant. This is not unexpected, because the total number of satellites is predicted using another correlation including the mass of the bulge through the bulge-to-total ratio. What came as a surprise is the sign of the correlation, because bulges are assumed to form through accretion of or merger with the surrounding material, including satellites. It would be natural to assume that the larger the bulge, the more satellites have been accreted, but we see quite the opposite: the larger the bulge, the more satellites have survived. 
This reasoning is valid only if we assume that galaxies with similar baryonic mass evolving in the similar sparse environment have had initially comparable satellite populations. 

A correlation that has been found in this work is the one between the total (expected) number of satellites and the total specific angular momentum. Angular momentum is the fundamental galactic property that is preserved during galaxy evolution. It can be redistributed via mergers, but preserved in general. SAGA galaxies with lower angular momentum have more satellites. Dynamical friction that can explain minor mergers would show the opposite behavior. The satellites should transfer their angular momentum to the host galaxy, slowing down and eventually spiraling in toward the central galaxy. The angular momentum of the host would consequently increase, because energy and momentum must be conserved.  This scenario would imply that galaxies with more massive bulges should harbor less satellites and have higher angular momentum. Results from this and previous work \citep{Javanmardi2019, Javanmardi2020} suggest that the more abundant satellite populations have more massive bulges that are not built up by minor mergers. If the satellites are formed within the disk of baryonic and dark matter around galaxies, carrying away part of its angular momentum, then this would provide a natural explanation of the plane of satellites observed in several nearby galaxies \citep{Pawlowski2013, Muller}. Deeper spectroscopic observations are needed to confirm these correlations because they critically depend on the number of satellites fainter than Leo I of the Milky Way. 


\begin{acknowledgements}
We acknowledge the financial support of the Ministry of Education, Science and Technological Development of the Republic of Serbia through the contract No.~451-03-68/2022-14/200002. The authors thank the referee for the prompt report and constructive comments. This research made use of data from the SAGA Survey (sagasurvey.org). The SAGA Survey was supported by NSF collaborative grants AST-1517148 and AST-1517422 and by Heising–Simons Foundation grant 2019-1402. This research made use of {\sc IRAF} and {\sc R} programming languages. We acknowledge the use of the Extragalactic Distance Database (\url{https://edd.ifa.hawaii.edu}). We acknowledge the usage of the HyperLeda database (\url{http://leda.univ-lyon1.fr}). Funding for the Sloan Digital Sky 
Survey IV has been provided by the Alfred P. Sloan Foundation, the U.S. Department of Energy Office of Science, and the Participating Institutions. 

SDSS-IV acknowledges support and resources from the Center for High Performance Computing  at the University of Utah. The SDSS website is www.sdss.org.

SDSS-IV is managed by the Astrophysical Research Consortium for the Participating Institutions of the SDSS Collaboration including the Brazilian Participation Group, the Carnegie Institution for Science, Carnegie Mellon University, Center for Astrophysics | Harvard \& Smithsonian, the Chilean Participation Group, the French Participation Group, Instituto de Astrof\'isica de Canarias, The Johns Hopkins University, Kavli Institute for the Physics and Mathematics of the 
Universe (IPMU) / University of Tokyo, the Korean Participation Group, Lawrence Berkeley National Laboratory, Leibniz Institut f\"ur Astrophysik Potsdam (AIP),  Max-Planck-Institut f\"ur Astronomie (MPIA Heidelberg), Max-Planck-Institut f\"ur 
Astrophysik (MPA Garching), Max-Planck-Institut f\"ur Extraterrestrische Physik (MPE), National Astronomical Observatories of China, New Mexico State University, 
New York University, University of Notre Dame, Observat\'ario Nacional / MCTI, The Ohio State University, Pennsylvania State University, Shanghai Astronomical Observatory, United Kingdom Participation Group, Universidad Nacional Aut\'onoma 
de M\'exico, University of Arizona, University of Colorado Boulder, University of Oxford, University of Portsmouth, University of Utah, University of Virginia, University of Washington, University of Wisconsin, Vanderbilt University, 
and Yale University.
      
\end{acknowledgements}
\onecolumn

\begin{table}[ht]
\caption{\label{tab:alldata}Properties of SAGA galaxies.}
\centering
\begin{tabular}{ccccccccccccc}
  \hline \hline
Name & $\log \mathcal{M}_{\rm bar}$ & $\log\mathcal{M}_*$ & $\log e\_\mathcal{M}$ & V & $e\_$V & $R_{\rm [3.6]}$ & $\log \mathcal{M}_{\rm dyn}$ & $\log e\_ \mathcal{M}_{\rm dyn}$ & $\log$ J & $\log$ e\_J & <N> & <e\_N> \\  
  \hline \hline
  1015 & 10.58 & 10.44 & 0.23 & 124.45 & 4.41 & 14.41 & 11.01 & 0.03 & 3.03 & 0.06 & 9.68 & 2.84 \\ 
  1309 & 10.60 & 10.49 & 0.23 & 94.33 & 5.50 & 11.91 & -- & -- & -- & -- & 3.80 & 1.71 \\ 
  2543 & 10.65 & 10.58 & 0.11 & 158.98 & 3.72 & 18.22 & 11.14 & 0.02 & 2.65 & 0.04 & 11.51 & 3.20 \\ 
  2962 & 10.71 & 10.69 & 0.11 & 117.70 & 5.89 & 19.90 & -- & -- & -- & -- & 10.59 & 3.02 \\ 
  2967 & 10.66 & 10.55 & 0.23 & 123.82 & 6.93 & 11.70 & 11.00 & 0.05 & 2.93 & 0.01 & 1.95 & 1.35 \\ 
  3689 & 10.59 & 10.56 & 0.15 & 176.51 & 9.18 & 11.98 & 10.96 & 0.05 & 3.17 & 0.01 & 3.34 & 1.62 \\ 
  3976 & 11.07 & 11.01 & 0.01 & 212.34 & 4.01 & 24.00 & 11.54 & 0.02 & 2.87 & 0.05 & 20.41 & 4.92 \\ 
  4158 & 10.67 & 10.64 & 0.23 & 161.36 & 9.63 & 10.16 & 10.93 & 0.05 & 2.27 & 0.52 & 28.17 & 6.42 \\ 
  4454 & 10.31 & 10.23 & 0.23 & 219.05 & 21.06 & 15.51 & -- & -- & -- & -- & 10.40 & 2.98 \\ 
  5297 & 10.81 & 10.62 & 0.23 & 189.47 & 5.03 & 19.63 & 11.55 & 0.02 & 3.10 & 0.03 & 5.17 & 1.97 \\ 
  5347 & 10.26 & 10.14 & 0.23 & 58.11 & 4.43 & 9.73 & -- & -- & -- & -- & 7.07 & 2.34 \\ 
  5448 & 10.70 & 10.63 & 0.02 & 198.49 & 6.08 & 20.26 & 11.35 & 0.03 & 3.14 & 0.01 & 10.39 & 2.98 \\ 
  5602 & 10.33 & 10.28 & 0.23 & -- & -- & 11.25 & -- & -- &-- & -- & 12.09 & 3.31 \\ 
  5604 & 10.49 & 10.31 & 0.21 & 163.29 & 4.20 & 10.77 & 11.25 & 0.02 & 2.94 & 0.01 & 2.18 & 1.39 \\ 
  5633 & 10.62 & 10.58 & 0.57 & 155.39 & 6.86 & 9.79 & 10.97 & 0.04 & 2.72 & 0.17 & 12.82 & 3.45 \\ 
  5690 & 10.50 & 10.43 & 0.26 & 146.89 & 4.01 & 15.42 & 11.00 & 0.02 & 2.48 & 0.24 & 19.50 & 4.74 \\ 
  5750 & 10.41 & 10.40 & 0.48 & 191.23 & 9.23 & 15.40 & 10.67 & 0.04 & 2.95 & 0.11 & 11.34 & 3.16 \\ 
  5792 & 10.87 & 10.75 & 0.09 & 214.30 & 5.08 & 24.35 & 11.60 & 0.02 & -- & -- & 15.85 & 4.04 \\ 
  5869 & 10.61 & 10.61 & 0.14 & 105.80 & 5.29 & 14.83 & -- & -- & -- & -- & 23.77 & 5.57 \\ 
  5962 & 10.58 & 10.54 & 0.31 & 188.42 & 4.46 & 12.98 & 11.12 & 0.02 & 2.89 & 0.11 & 12.95 & 3.48 \\ 
  6181 & 10.73 & 10.66 & 0.01 & 192.47 & 4.25 & 12.17 & 11.32 & 0.02 & 2.75 & 0.02 & 12.17 & 3.33 \\ 
  6278 & 10.67 & 10.64 & 0.23 & 189.80 & 9.49 & 12.47 & 11.15 & 0.04 & 2.80 & 0.13 & 15.80 & 4.03 \\ 
  6909 & 10.55 & 10.52 & 0.01 & -- & -- & 11.25 & -- & -- &-- & -- & 15.61 & 3.99 \\ 
  7029 & 10.79 & 10.77 & 0.01 & -- & -- & 11.25 & -- & -- &-- & -- & 12.22 & 3.33 \\ 
  7079 & 10.71 & 10.71 & 0.30 & 229.09 & 10.99 & 14.27 & 10.88 & 0.04 & 3.23 & 0.06 & 10.03 & 2.91 \\ 
  7328 & 10.41 & 10.32 & 0.02 & 157.45 & 3.06 & 12.21 & 11.06 & 0.02 & -- &  -- & 6.74 & 2.28 \\ 
  7541 & 10.94 & 10.85 & 0.17 & 211.56 & 3.01 & 23.54 & 11.58 & 0.01 & 3.18 & 0.01 & 2.28 & 1.41 \\ 
  7716 & 10.44 & 10.35 & 0.23 & 148.96 & 4.53 & 11.39 & 11.02 & 0.03 & 2.71 & 0.05 & 9.85 & 2.88 \\ 
  68743 & 10.49 & 10.44 & 0.15 & 43.51 & 2.43 & 12.72 & -- & -- & -- & -- & 17.22 & 4.30 \\ 
   \hline
\end{tabular}
\tablefoot{Column (1): NGC number, where NGC is omitted for all four-digit entries, and a single five-digit number refers to a PGC galaxy (the last entry). Columns (2-4): baryonic and stellar masses (in the solar masses) with the corresponding error (the same for both). Columns (5-6): maximum rotation velocity with the corresponding error in units of $\rm{km\ s^{-1}}$. Column (7): radius at which galaxy surface brightness profile falls to 25.5 mag/arcsec$^2$ in units of kpc. Columns (8-9): dynamical mass and the corresponding error in the solar masses. Columns (9-11): total specific angular momentum with the corresponding error in units of ${\rm km\ s^{-1}}$ kpc. Columns (12-13): expected number of satellites with an error as a square root of the number of satellites. }
\end{table}

%
%
\bibliographystyle{aa}
\bibliography{saga}

%
%

%

%
%

%


\begin{appendix} 
\section{Table of structural and kinematic properties of SAGA galaxies}
\label{app_table}
Multicomponent modeling of SAGA galaxies with {\sc GALFIT} code, from which the mass of the bulge is estimated, as well as the structural parameters of the bulge: Sersic index n, effective radius $R_{\rm eff}$ and ellipticity $\epsilon=1-b/a$. In addition, velocity dispersions are retrieved from: the  Sloan Digital Sky Survey catalog and the Hyperleda database. Radius of HI disk is estimated using Eq.~\ref{eq:RHI} from gas mass $M_{\rm HI}$ that was retrieved from the Extragalactic Distance Database.
\begin{table}[ht]
\caption{\label{tab:all_data_app} Structural and kinematic properties of SAGA galaxies}
\centering
\begin{tabular}{cccccccccccccc}
  \hline \hline
NGC & $\mathcal{M}_{\rm b}$ & e\_$\mathcal{M}_{\rm b}$ & n & R$_{\rm eff}$ & $\epsilon$ & $\sigma_0$ & e\_$\sigma_0$ & $\mathcal{M}_{\rm d}$ & e\_$\mathcal{M}_{\rm d}$ & R$_{\rm h}$ & R$_{\rm HI}$ & B/T$_{*}$ & e\_B/T$_{*}$\\ 
& [$10^{10}$ M$_{\odot}$] & [$10^{10}$ M$_{\odot}$] & & [kpc] & & [km/s] & [km/s] & [$10^{10}$ M$_{\odot}$] & [$10^{10}$ M$_{\odot}$] & [kpc] & [kpc] & & \\
  \hline \hline
  1015 & 0.90 & 0.02 & 1.00 & 11.17 & 0.07 & 106.50 & 6.60 & 1.83 & 0.03 & 6.34 & 28.72 & 0.33 & 0.01 \\ 
  1309 & 0.25 & 0.01 & 2.72 & 11.29 & 0.12 & 82.00 & 26.70 & 2.85 & 0.05 & 2.94 & 26.86 & 0.08 & 0.01 \\ 
  2543 & 1.31 & 0.03 & 1.00 & 2.87 & 0.41 & 79.43 & 3.60 & 2.47 & 0.04 & 2.10 & 23.41 & 0.35 & 0.01 \\ 
  2962 & 1.35 & 0.03 & 2.35 & 4.99 & 0.42 & 171.00 & 8.60 & 3.57 & 0.13 & 6.49 & 11.60 & 0.27 & 0.01 \\ 
  2967 & 0.00 & -- & -- & -- & 0.07 & 117.30 & 29.60 & 3.48 & 0.06 & 3.46 & 28.08 & 0.01 & 0.01 \\ 
  3689 & 0.00 & -- & -- & -- & 0.38 & -- & -- & 3.45 & 0.06 & 4.22 & 12.55 & 0.05 & 0.01 \\ 
  3976 & 6.56 & 0.26 & 2.89 & 8.82 & 0.74 & 190.10 & 8.00 & 3.79 & 0.06 & 3.29 & 33.67 & 0.63 & 0.02 \\ 
  4158 & 3.66 & 0.18 & 1.00 & 16.33 & 0.25 & 92.73 & 2.00 & 0.72 & 0.01 & 2.22 & 14.21 & 0.84 & 0.04 \\ 
  4454 & 0.52 & 0.02 & 1.00 & 5.99 & 0.11 & -- & -- & 1.16 & 0.03 & 5.04 & 16.18 & 0.31 & 0.01 \\ 
  5297 & 0.67 & 0.02 & 5.48 & 7.27 & 0.81 & 118.87 & 20.31 & 3.47 & 0.08 & 3.74 & 42.78 & 0.16 & 0.01 \\ 
  5347 & 0.29 & 0.01 & 1.00 & 1.13 & 0.34 & 93.90 & 14.91 & 1.10 & 0.02 & 1.87 & 18.29 & 0.21 & 0.01 \\ 
  5448 & 1.29 & 0.03 & 3.31 & 6.33 & 0.59 & 133.66 & 2.75 & 2.93 & 0.07 & 4.88 & 24.55 & 0.31 & 0.01 \\ 
  5602 & 0.66 & 0.01 & 2.19 & 3.52 & 0.46 & -- & -- & 1.25 & 0.02 & 2.34 & 13.31 & 0.35 & 0.01 \\ 
  5604 & 0.04 & 0.00 & 1.00 & 18.56 & 0.44 & -- & -- & 2.00 & 0.10 & 2.69 & 28.63 & 0.02 & 0.00 \\ 
  5633 & 1.37 & 0.03 & 0.17 & 20.46 & 0.30 & 73.31 & 4.09 & 2.41 & 0.04 & 2.51 & 16.60 & 0.36 & 0.01 \\ 
  5690 & 1.69 & 0.04 & 1.00 & 45.95 & 0.71 & 40.15 & 6.34 & 0.98 & 0.01 & 1.67 & 20.12 & 0.63 & 0.02 \\ 
  5750 & 0.73 & 0.01 & 2.81 & 23.47 & 0.52 & 95.06 & 6.13 & 1.78 & 0.03 & 3.10 & 5.49 & 0.29 & 0.01 \\ 
  5792 & 3.09 & 0.11 & 1.00 & 38.56 & 0.57 & -- & -- & 2.47 & 0.04 & 2.52 & 37.54 & 0.56 & 0.02 \\ 
  5869 & 2.66 & 0.22 & 6.90 & 4.46 & 0.37 & 174.60 & 8.70 & 1.38 & 0.02 & 1.32 & 0.00 & 0.66 & 0.05 \\ 
  5962 & 1.27 & 0.07 & 2.99 & 16.80 & 0.33 & 103.32 & 2.71 & 2.18 & 0.04 & 3.06 & 16.07 & 0.37 & 0.02 \\ 
  6181 & 1.68 & 0.05 & 0.38 & 25.59 & 0.41 & 124.80 & 9.00 & 2.94 & 0.05 & 1.97 & 24.42 & 0.36 & 0.01 \\ 
  6278 & 1.98 & 0.05 & 2.45 & 2.99 & 0.30 & 202.80 & 10.10 & 2.35 & 0.26 & 2.97 & 16.95 & 0.46 & 0.03 \\ 
  6909 & 1.48 & 0.05 & 0.86 & 20.60 & 0.53 & 117.16 & 2.85 & 1.84 & 0.03 & 3.58 & 13.95 & 0.45 & 0.02 \\ 
  7029 & 1.94 & 0.04 & 2.74 & 11.98 & 0.41 & 185.01 & 7.16 & 3.97 & 0.08 & 3.33 & 14.95 & 0.33 & 0.01 \\ 
  7079 & 1.28 & 0.02 & 1.46 & 3.98 & 0.36 & 158.90 & 4.56 & 3.82 & 0.07 & 4.96 & 6.29 & 0.25 & 0.01 \\ 
  7328 & 0.39 & 0.03 & 3.55 & 8.81 & 0.55 & -- & -- & 1.69 & 0.03 & 2.80 & 19.89 & 0.19 & 0.01 \\ 
  7541 & 0.00 & 0.00 & -- & -- & 0.71 & 65.04 & 34.94 & 6.89 & 0.11 & 3.60 & 36.73 & 0.02 & 0.01 \\ 
  7716 & 0.67 & 0.01 & 1.95 & 3.74 & 0.20 & 115.33 & 2.39 & 1.57 & 0.03 & 2.41 & 20.43 & 0.30 & 0.01 \\ 
  68743 & 1.42 & 0.03 & 1.45 & 6.61 & 0.30 & 156.90 & 12.70 & 1.35 & 0.02 & 2.76 & 15.35 & 0.51 & 0.01 \\ 
   \hline
\end{tabular}
\tablefoot{Column (1): NGC number, where NGC is omitted for all four-digit entries, and a single five-digit number refers to PGC galaxy. Columns (2-3): mass of the bulge with its corresponding error in the units of $10^{10}$ solar masses. Column (4): Sersic index n of the bulge. Column (5): effective radius of the bulge in kpc. Column (6): ellipticity of the bulge. Columns (7-8): velocity dispersion and the corresponding error in $\rm {km\ s^{-1}}$. Columns (9-10): mass of the disk component with the error in the units of $10^{10}$ solar masses. Column (11): disk scale length in kpc. Column (12): radius of the HI disk. Columns (13-14): bulge-to-total ratio with the corresponding error.}
\end{table}

\section{Surface brightness profiles of SAGA galaxies - decomposition into bulge and disk}
\label{app_plots}
Near-infrared images from the Spitzer Heritage Archive in 3.6 and 4.5 $\mu$m were retrieved and modeled with concentric ellipses using the IRAF ellipse task. These radial profiles are given in Fig.~\ref{fig:radprofs}.1 for both bands. Isophotal points from the {\tt ellipse} task that deviate from the radial trend are artifacts from modeling without constraints on magnitude error and the number of valid points (unmasked pixels) inside ellipses. Constraints have not been imposed to recover the faintest parts of galaxies (the furthest points away from the center). They are not used for any analysis, and are given only as a representation of {\sc GALFIT} modeling results in two dimensions. Images are also modeled with {\sc GALFIT} code to extract individual components: bulge and disk. Apart from these two components, additional components were often required to obtain successful modeling: point spread function (PSF) for the nucleus and Sersic with low index $n$ for bars and rings. Individual components are given for the 3.6 $\mu$m band only, for clarity. Composite models are generated as the sum of the fluxes of individual components: black dash dotted line in 3.6 $\mu$m, and gray dash dotted
line in 4.5 $\mu$m. For comparison, isophotal contours from \citet{MunozMateos2013} are overplotted in the Fig.~\ref{fig:radprofs}.1. The results from this work agree well with the results from \citet{MunozMateos2013}. 

Three galaxies are edge-on and thus excluded from further study: NGC~4348, PGC~013646 and UGC~00903. Their radial profiles could not be modeled well with elliptical contours due to an extreme inclination. They are listed here only for completeness.

\begin{figure}[ht]
\subfloat{\includegraphics[width = .48 \textwidth]{images/NGC1015_sb_color.pdf}}
\subfloat{\includegraphics[width = .48\textwidth]{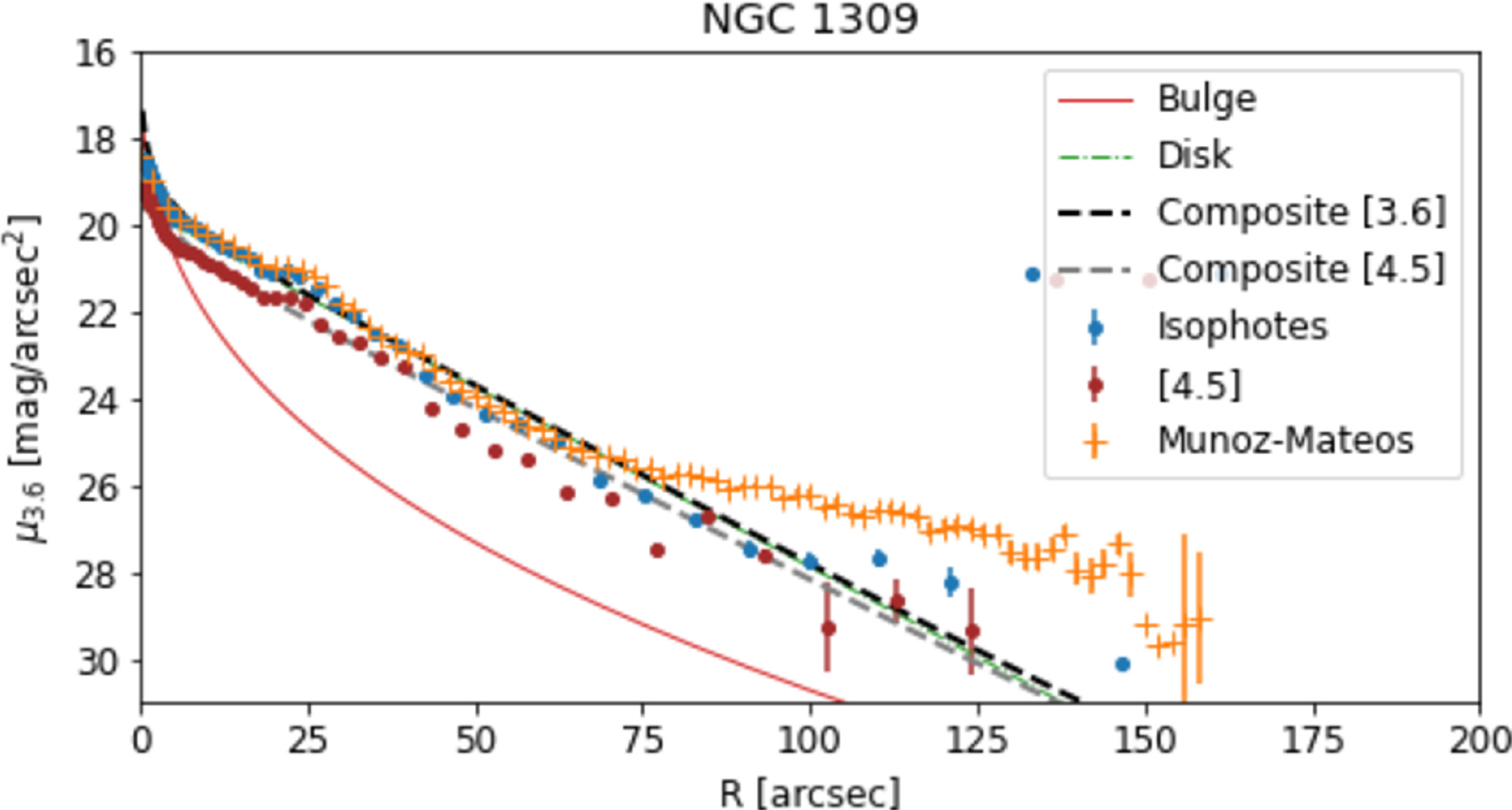}}\\ 
\subfloat{\includegraphics[width = .48\textwidth]{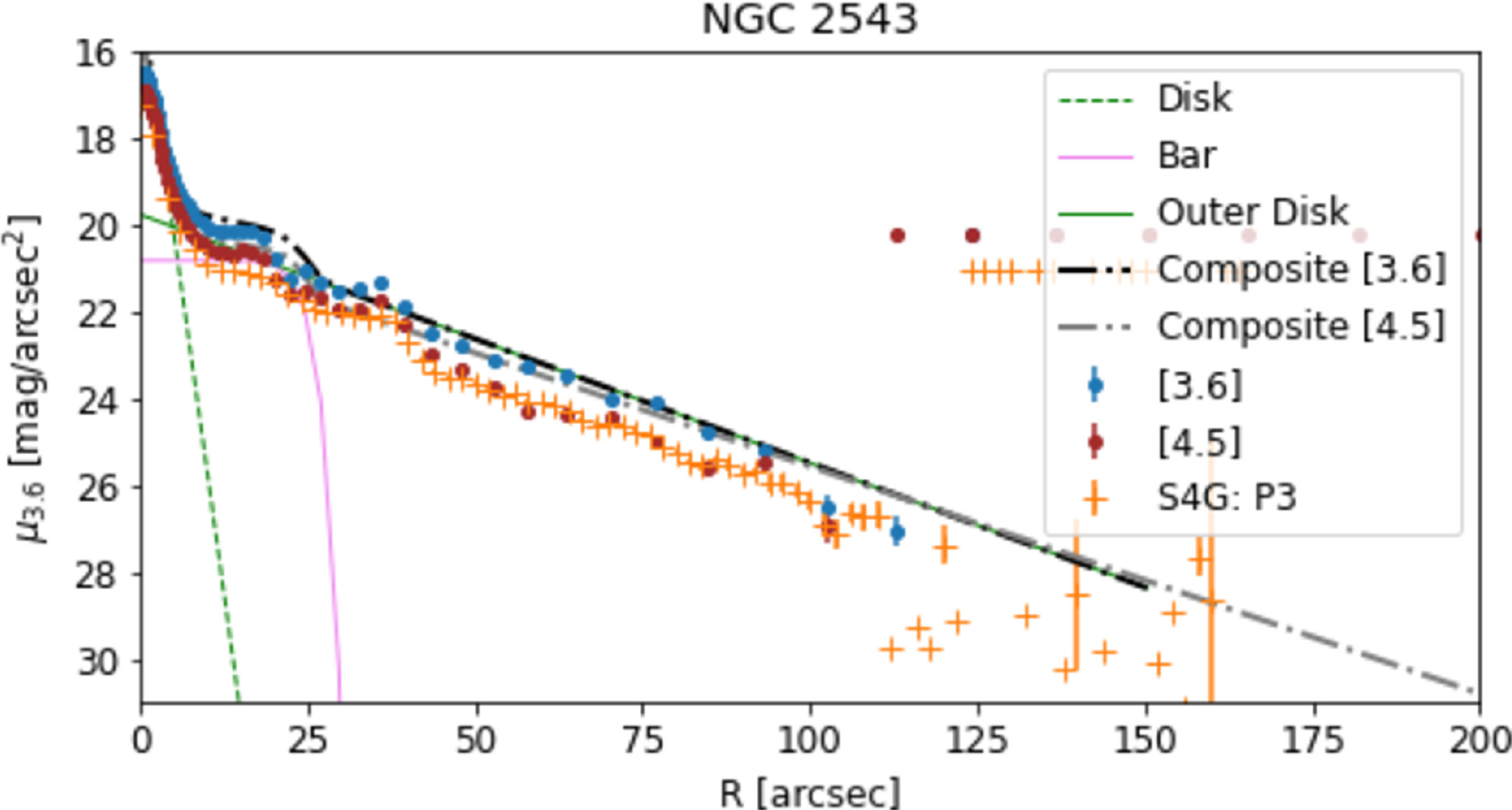}}
\subfloat{\includegraphics[width = .48\textwidth]{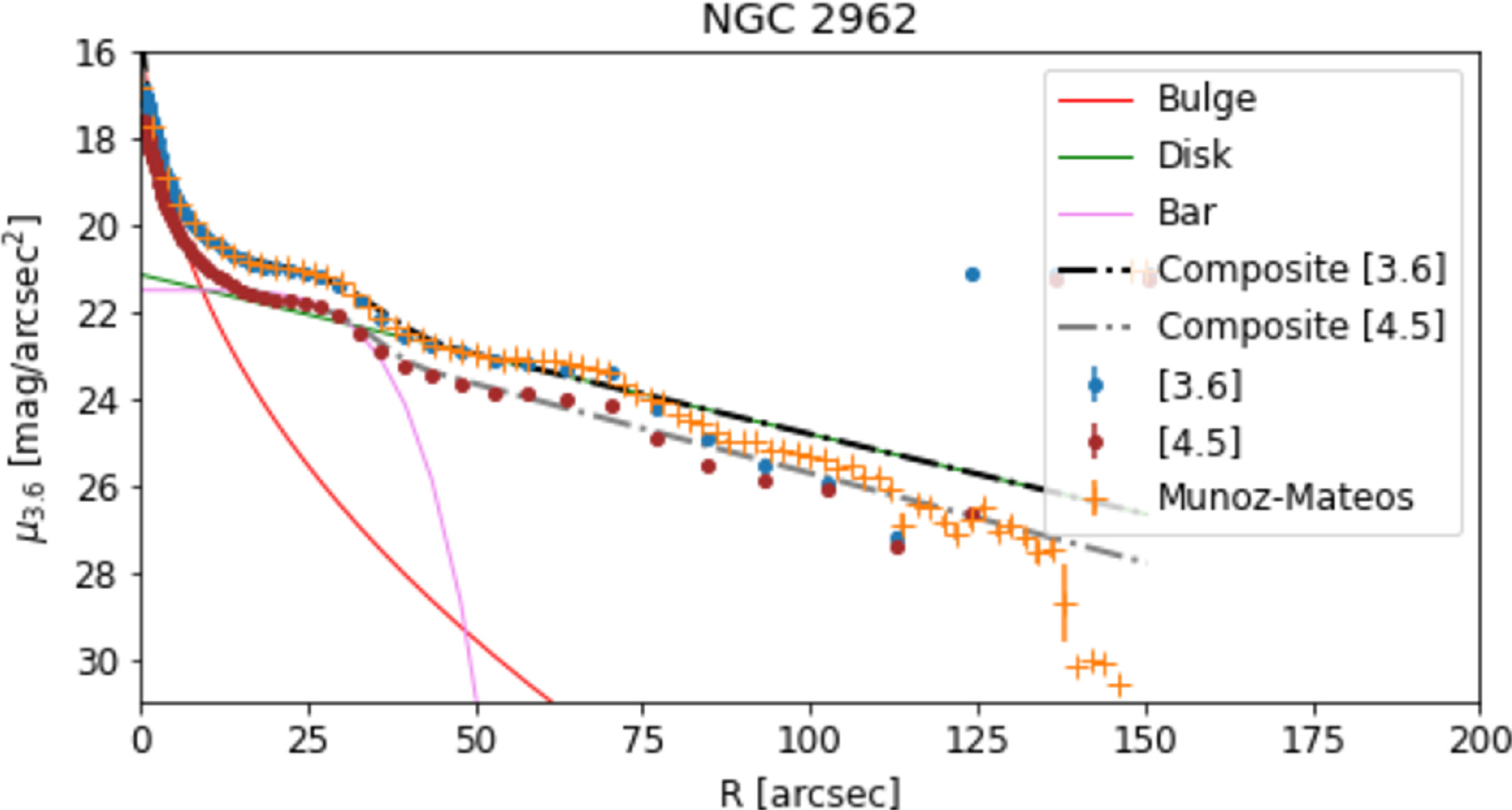}}\\ 
\caption{Surface brightness profiles of 32 SAGA galaxies in the NIR (3.6 and 4.5 $\mu$m). All galaxies are modeled with multiple components including bulge and disk, but also other components such as the nucleus, ring, and bar. Individual components are colored in the same manner in all the subfigures: PSF in orange, bulge in red, disk in green, and bar/ring in violet. Blue circles correspond to radial profile in 3.6 $\mu$m, while red circles correspond to 4.5 $\mu$m. Individual modeled components are presented for clarity only for 3.6 $\mu$m. The composite model is given for both 3.6 $\mu$m (black dash dotted line) and 4.5 $\mu$m (gray dash dotted line). The figure continues on the next page.}
\label{fig:radprofs}
\end{figure}
\begin{figure}
\ContinuedFloat 
\subfloat{\includegraphics[width = .48\textwidth]{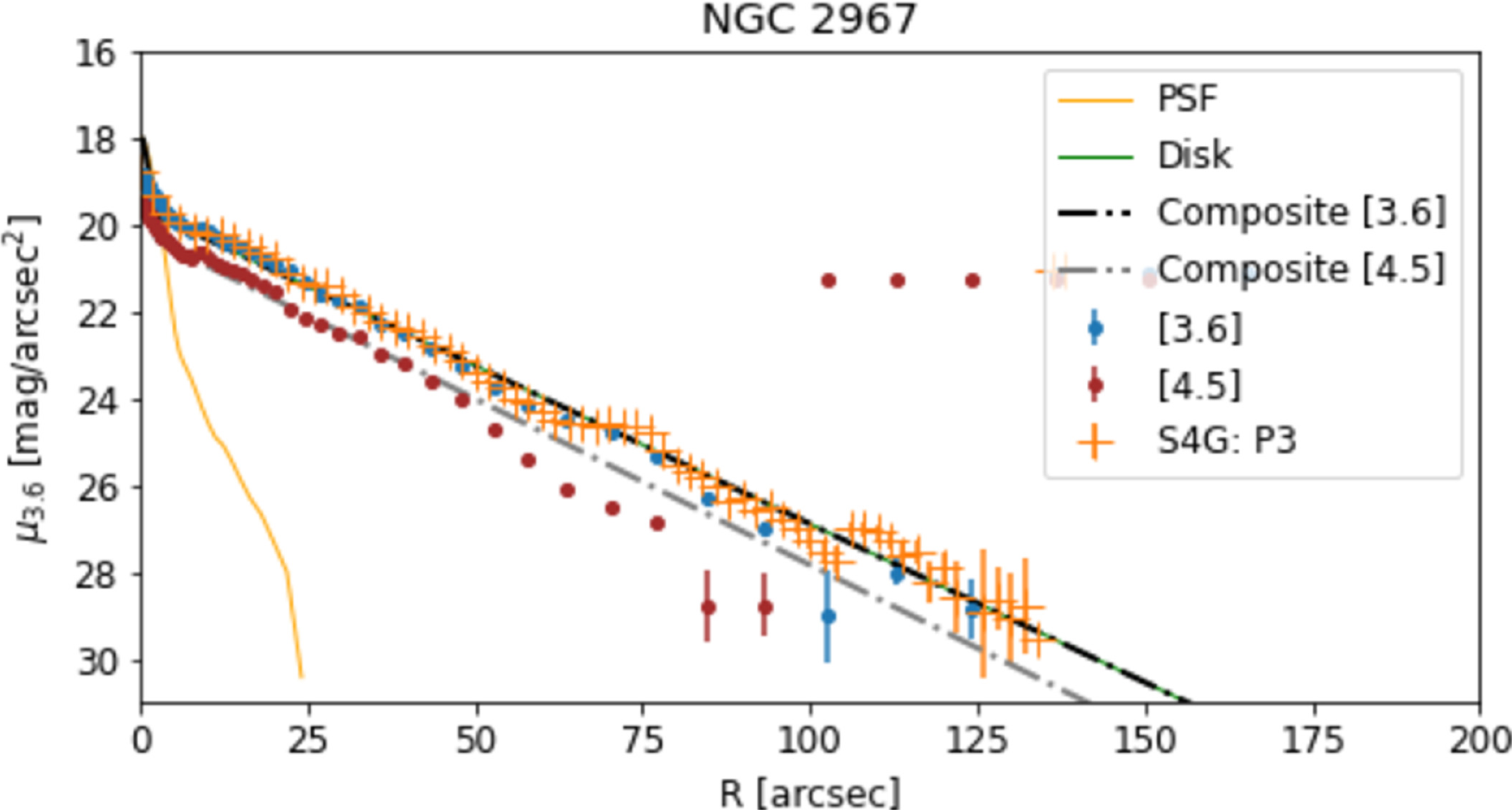}}
\subfloat{\includegraphics[width=0.48\textwidth]{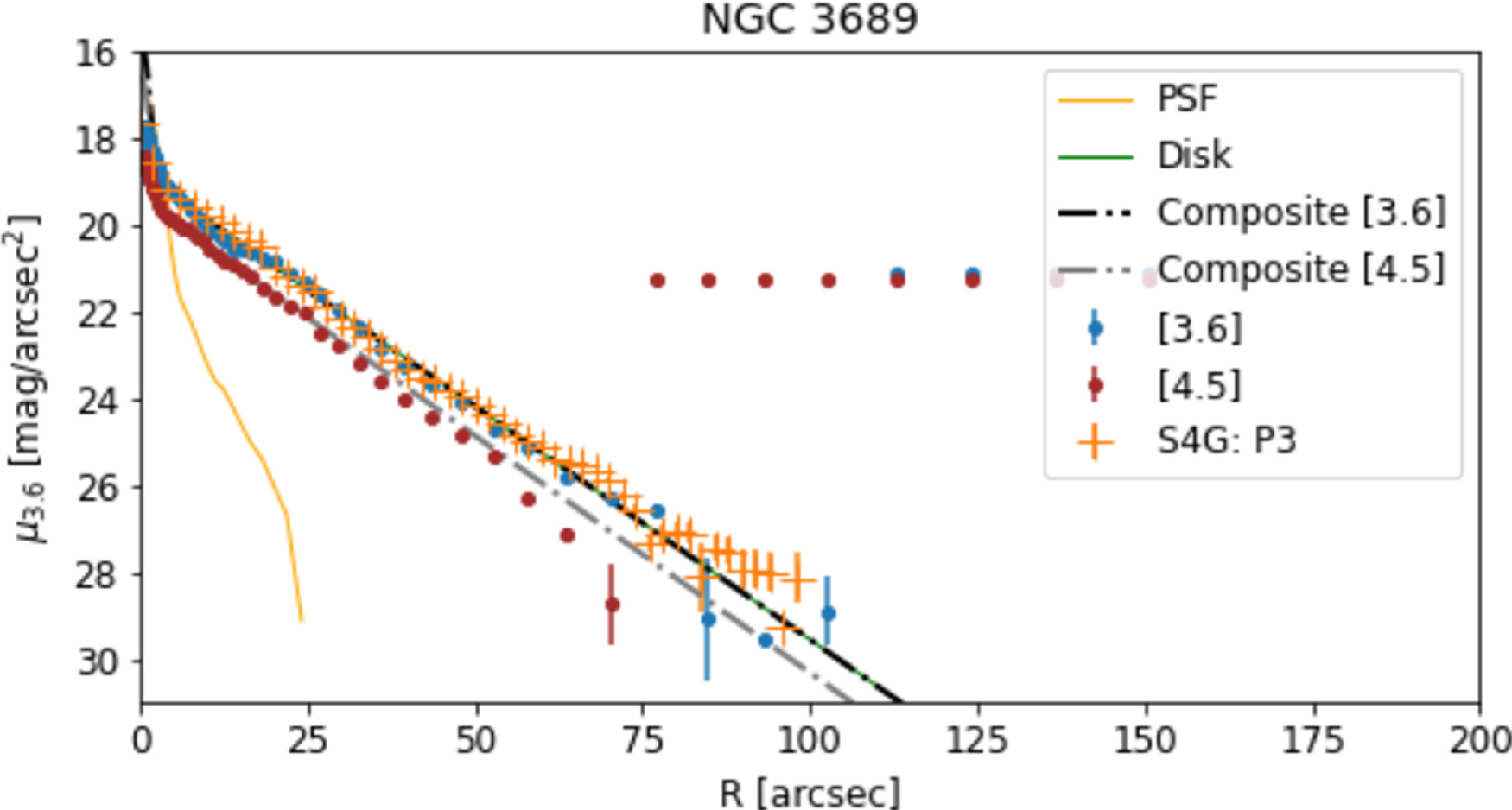}}\\ 
\subfloat{\includegraphics[width=0.48\textwidth]{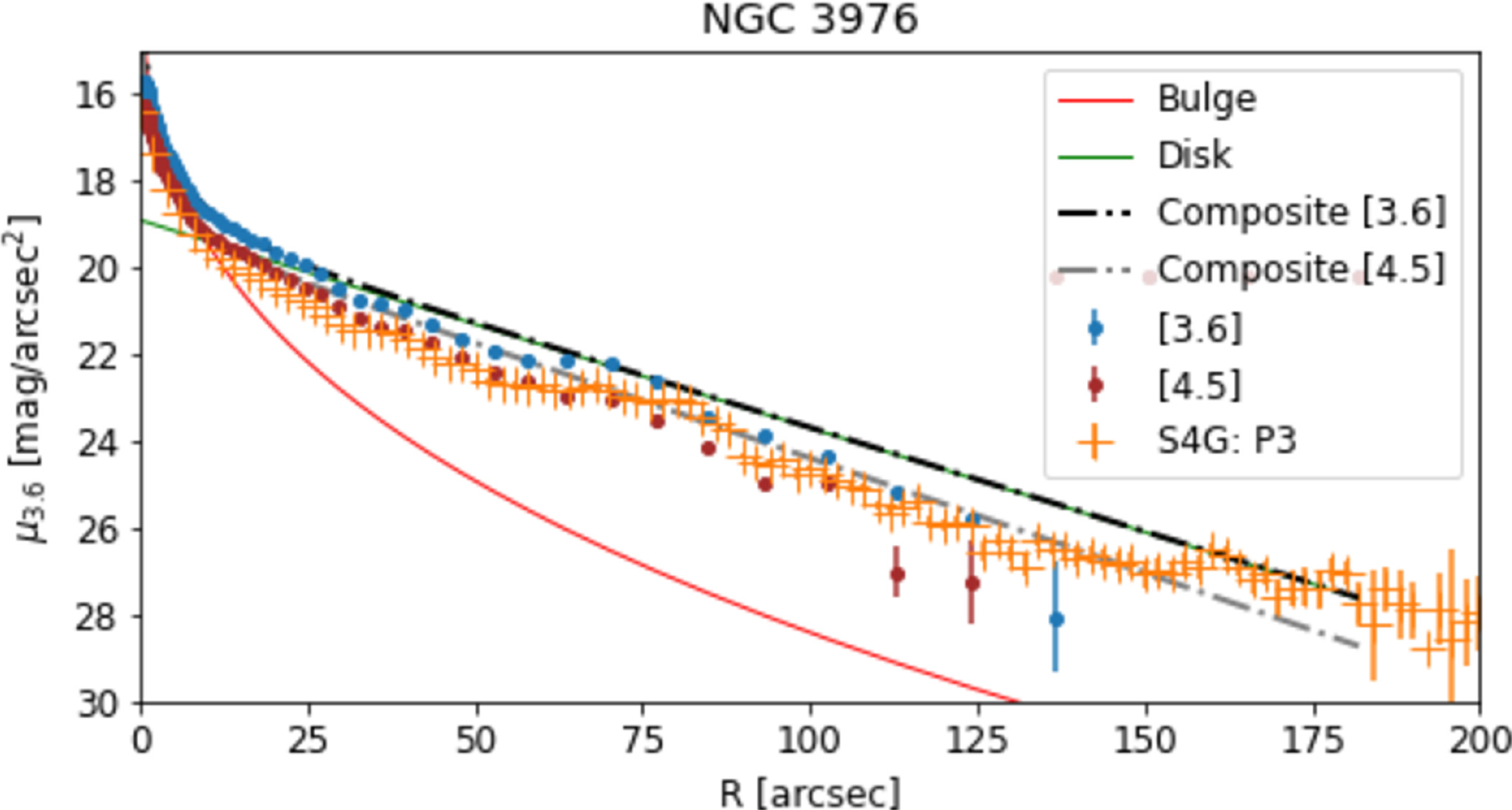}}
\subfloat{\includegraphics[width=0.48\textwidth]{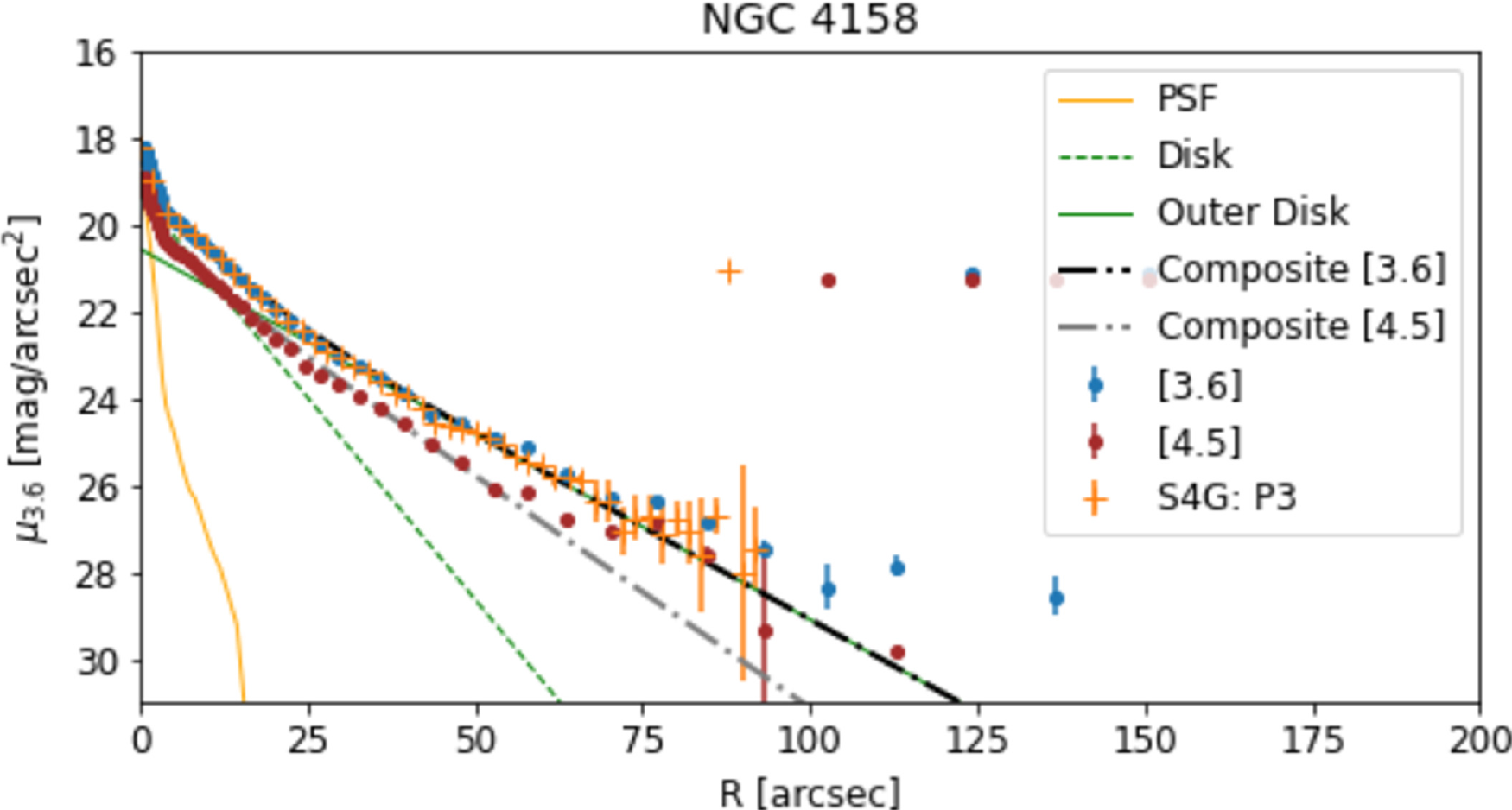}} \\
\subfloat{\includegraphics[width=0.48\textwidth]{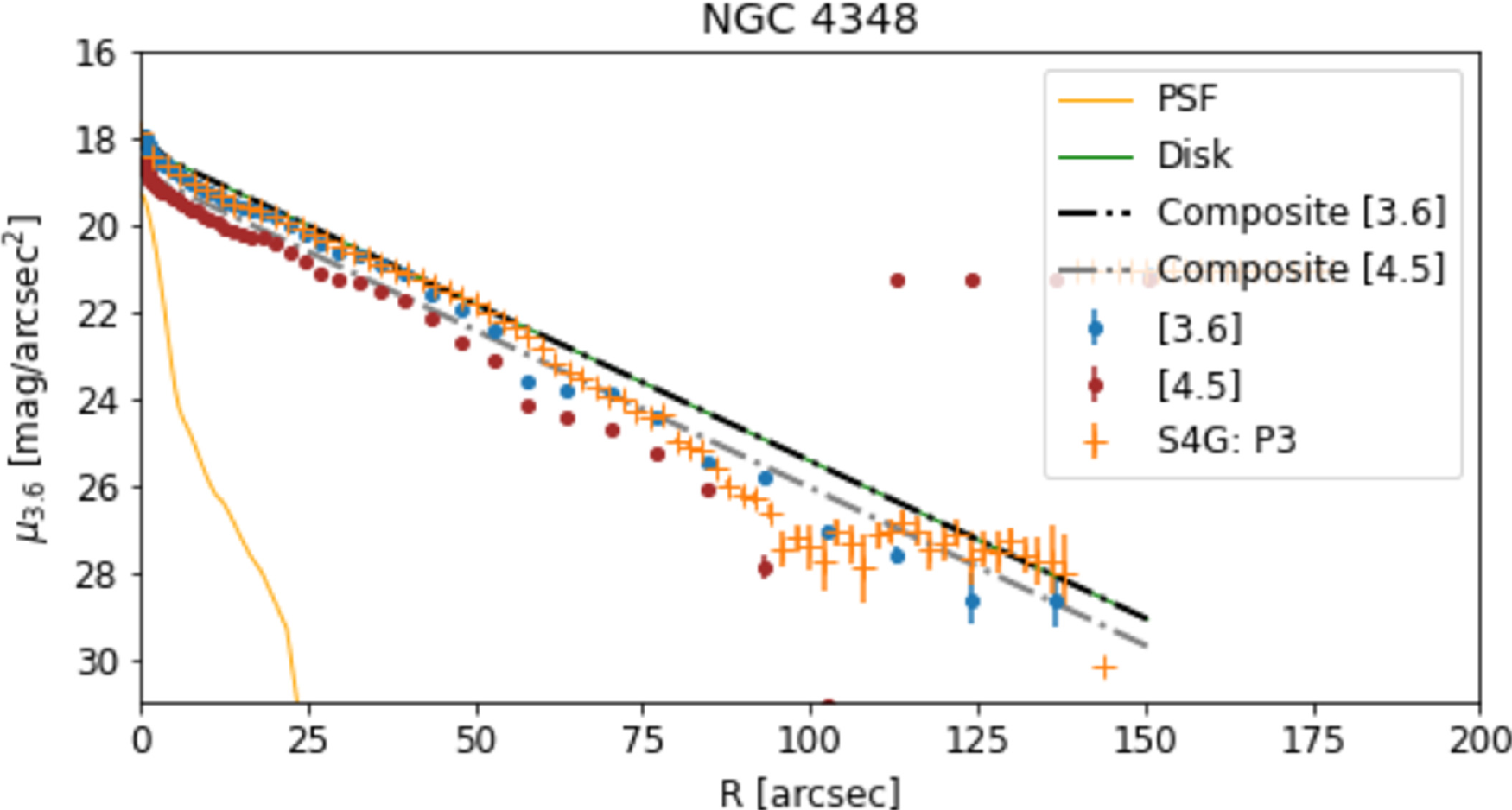}}
\subfloat{\includegraphics[width=0.48\textwidth]{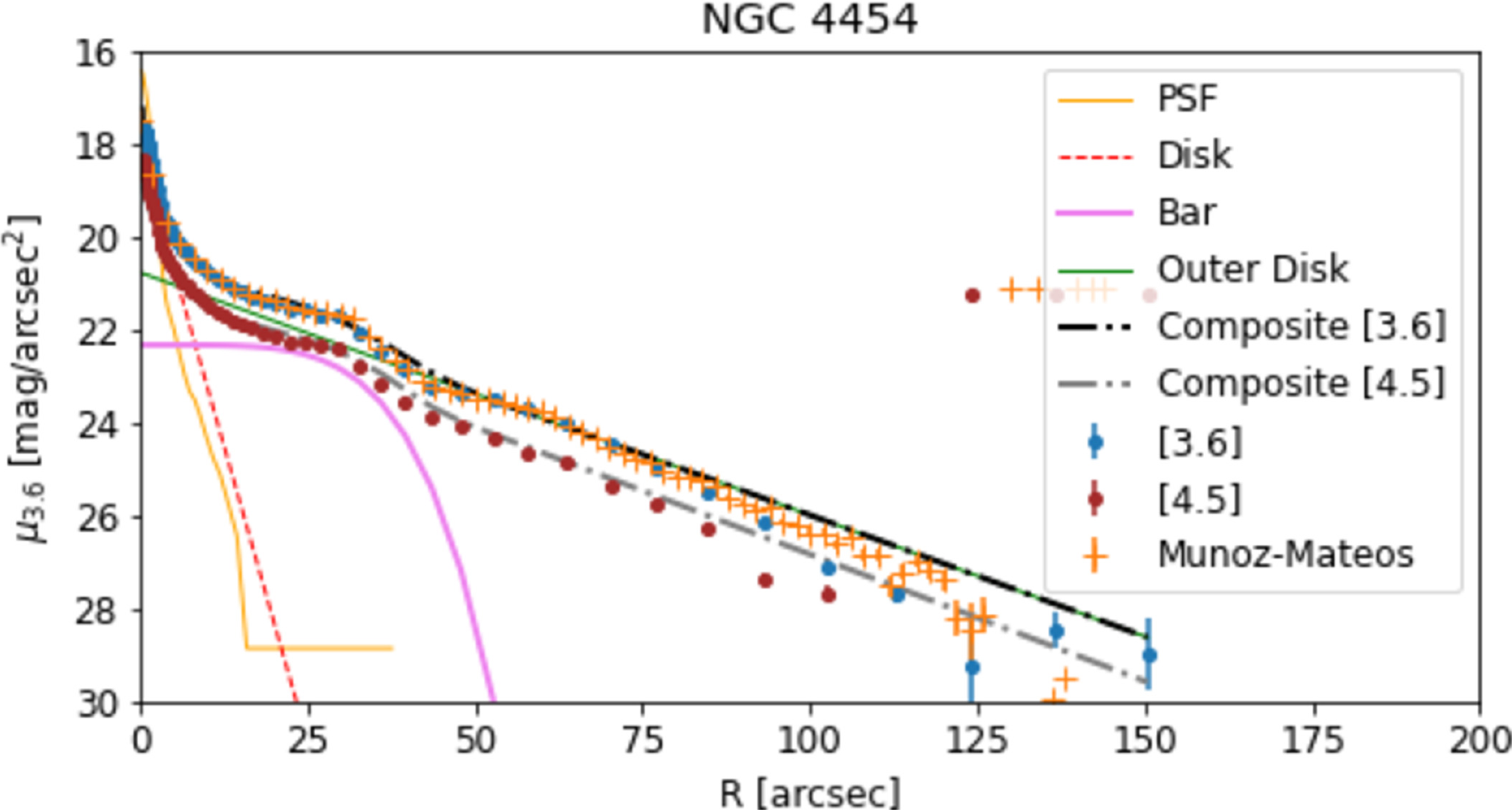}} \\
\subfloat{\includegraphics[width=0.48\textwidth]{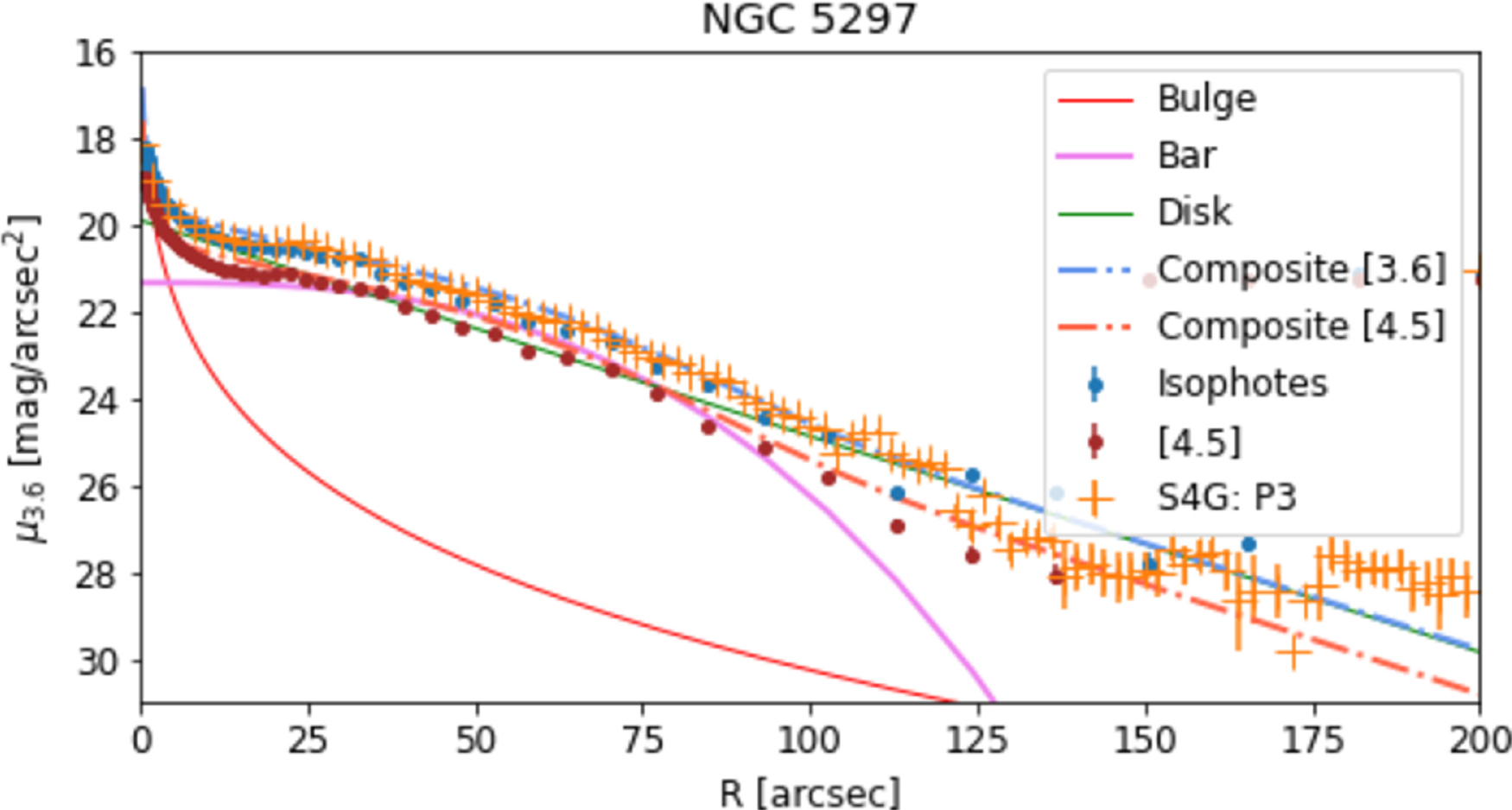}}
\subfloat{\includegraphics[width=0.48\textwidth]{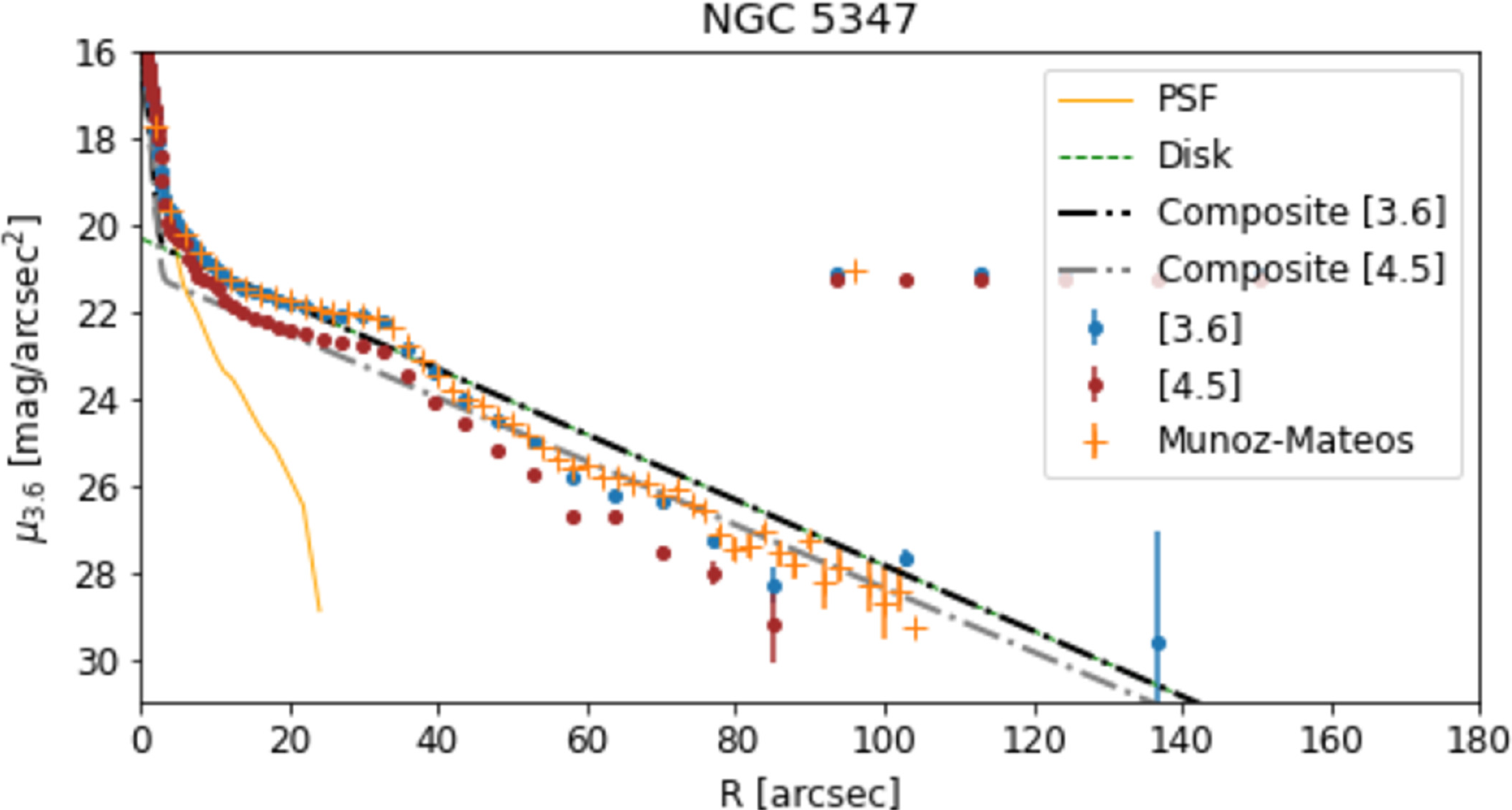}} \\
\subfloat{\includegraphics[width=0.48\textwidth]{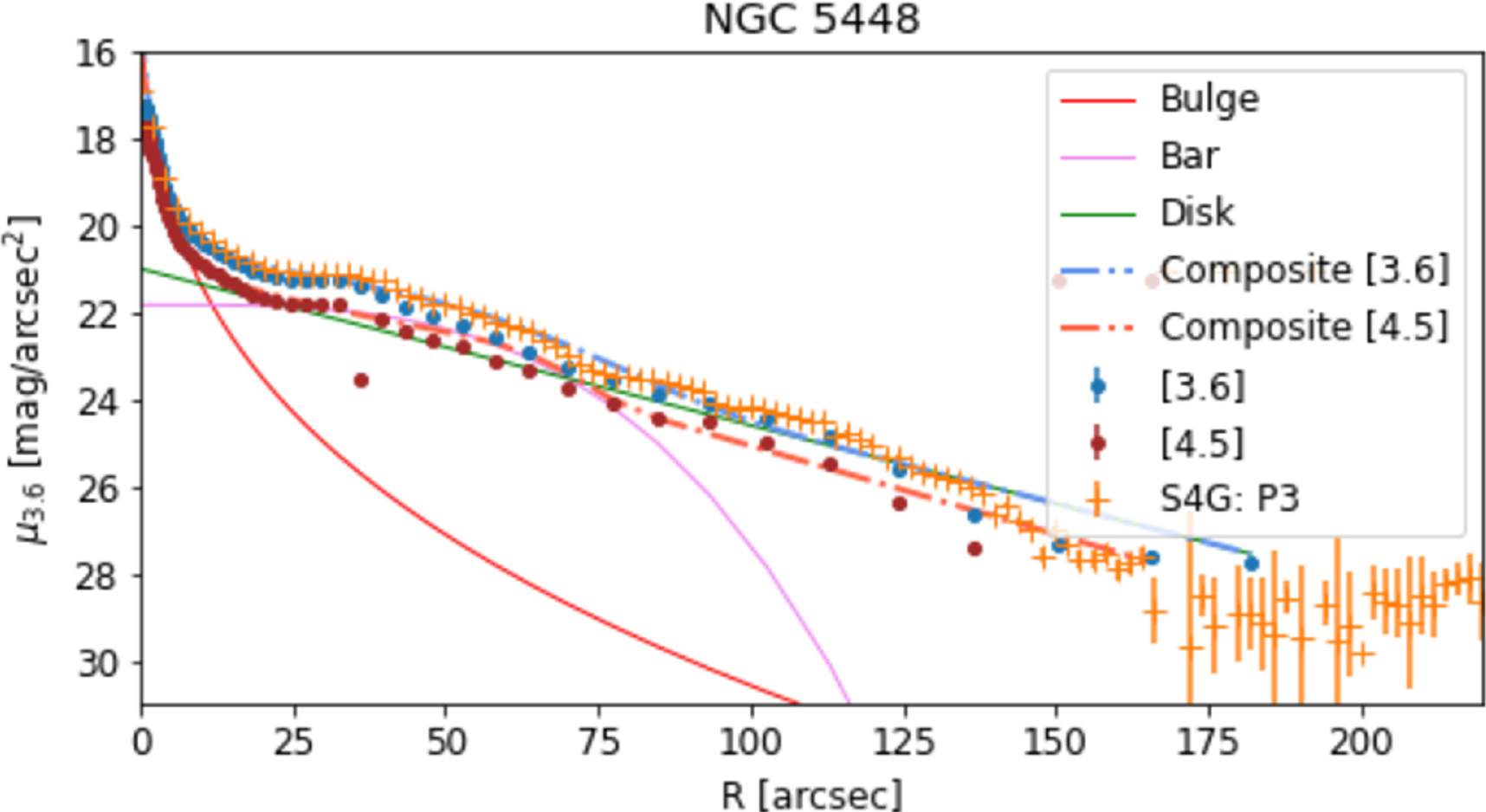}}
\subfloat{\includegraphics[width=0.48\textwidth]{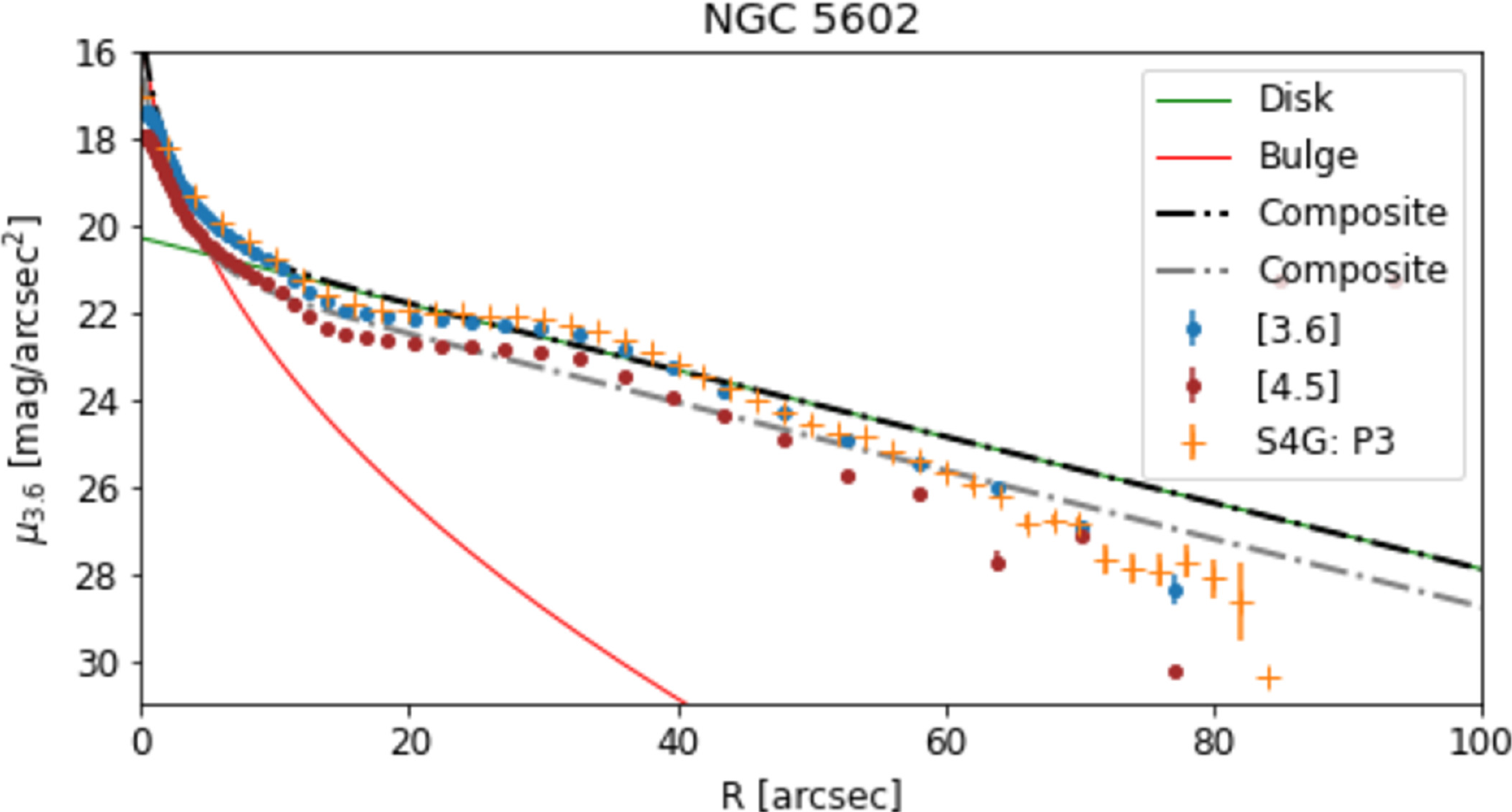}} \\
\caption[]{continued.}
\end{figure}
\begin{figure}
\ContinuedFloat 
\subfloat{\includegraphics[width=0.48\textwidth]{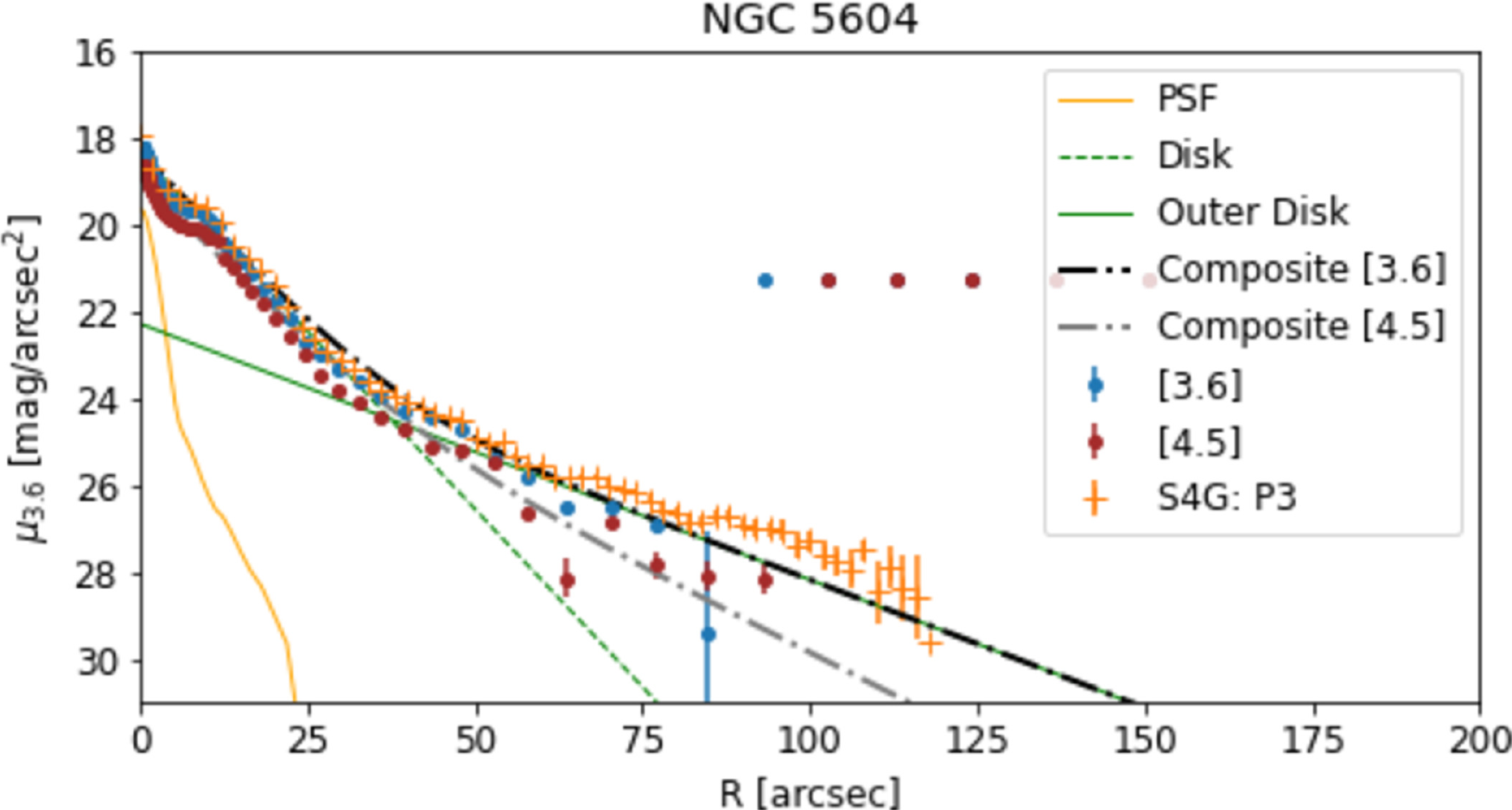}}
\subfloat{\includegraphics[width=0.48\textwidth]{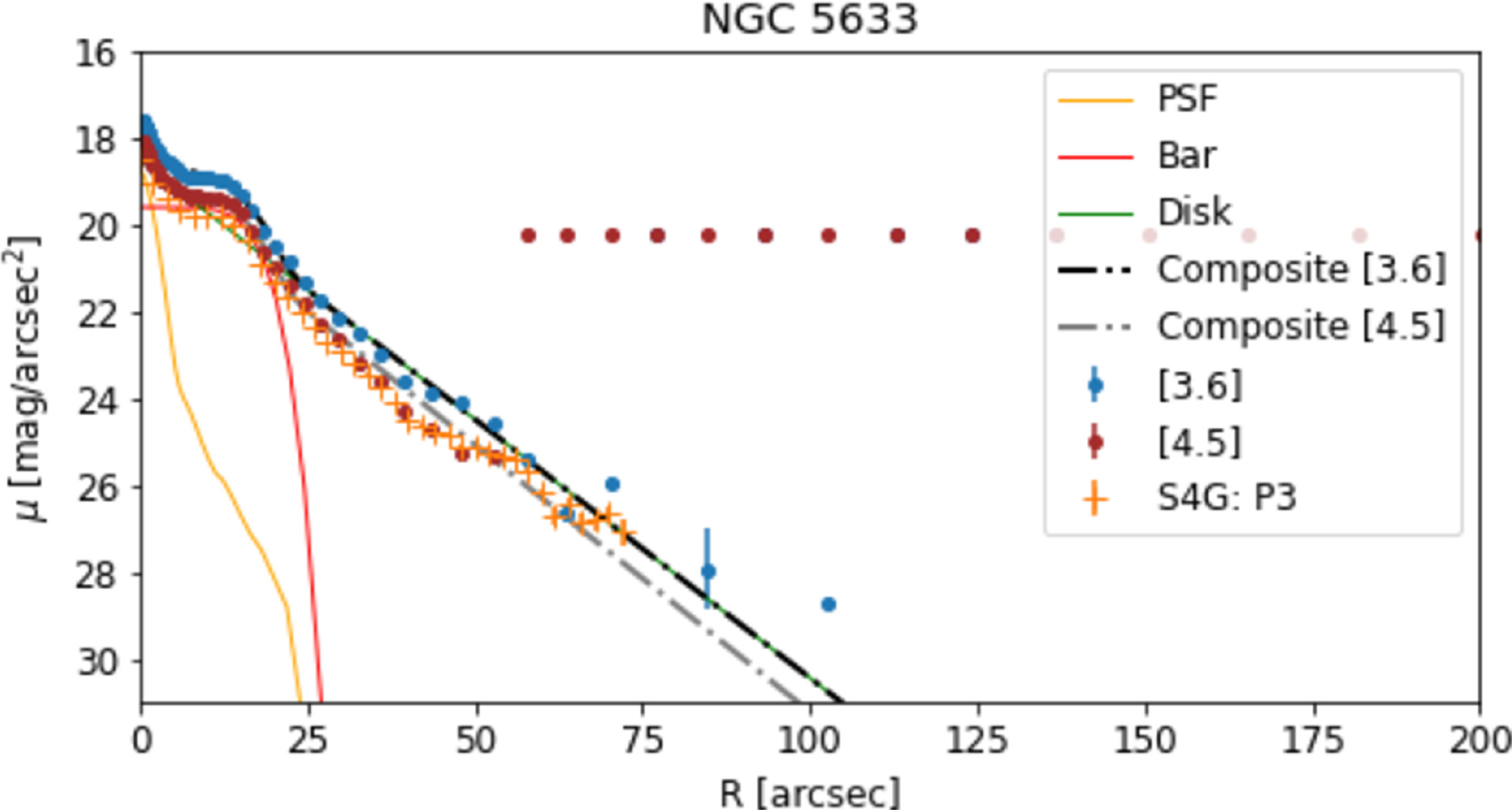}} \\
\subfloat{\includegraphics[width=0.48\textwidth]{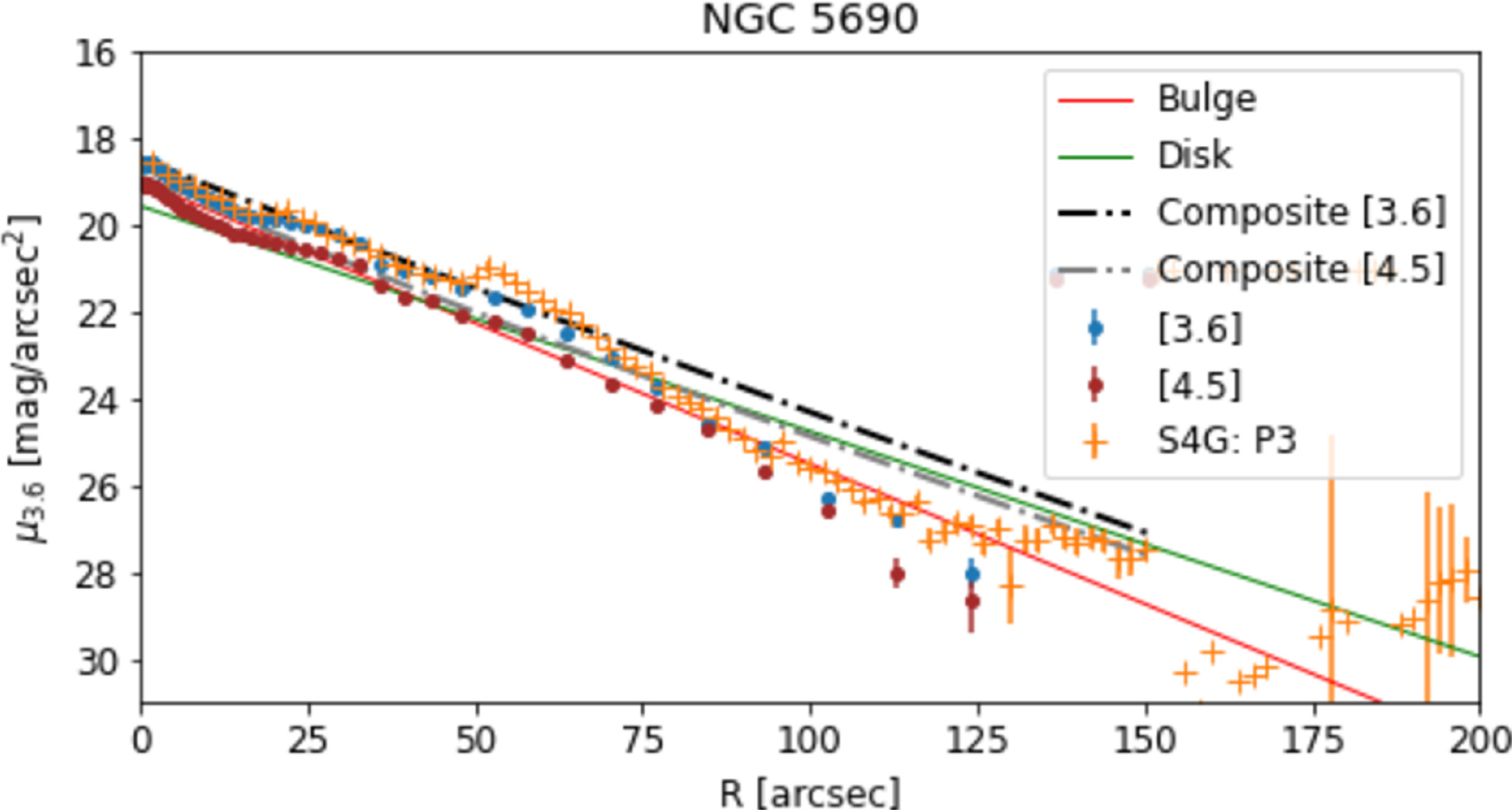}}
\subfloat{\includegraphics[width=0.48\textwidth]{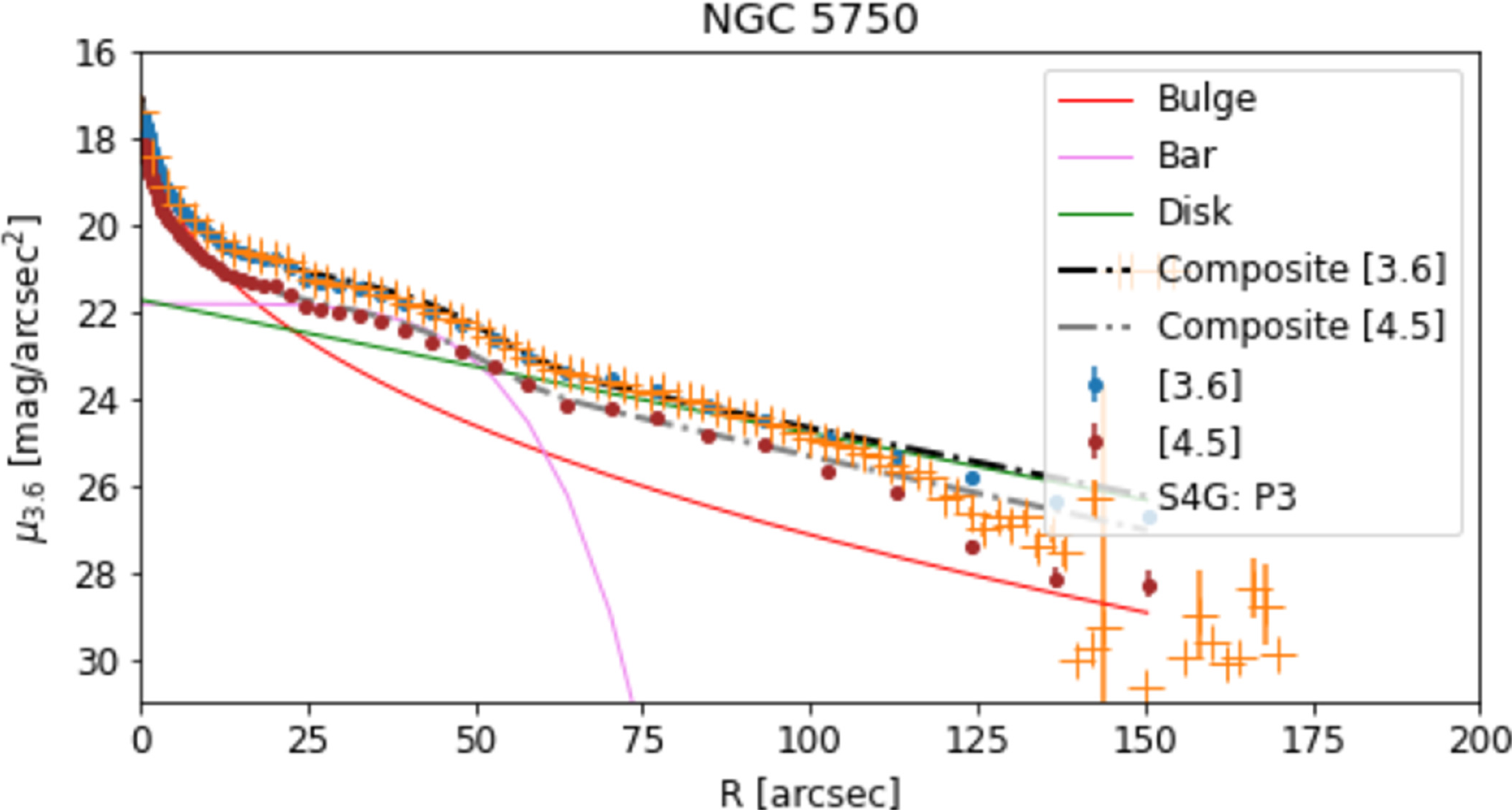}}   \\
\subfloat{\includegraphics[width=0.48\textwidth]{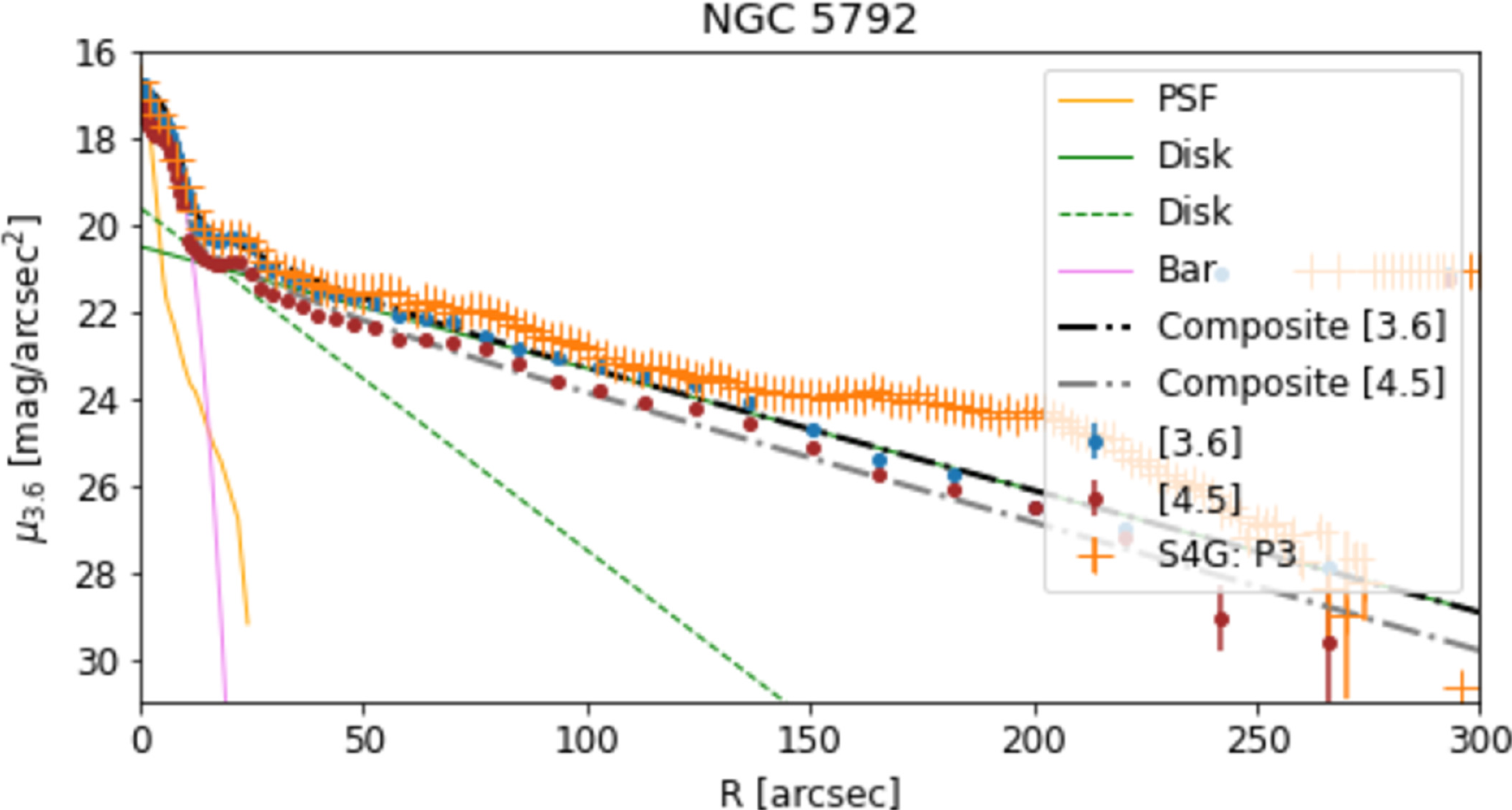}}
\subfloat{\includegraphics[width=0.48\textwidth]{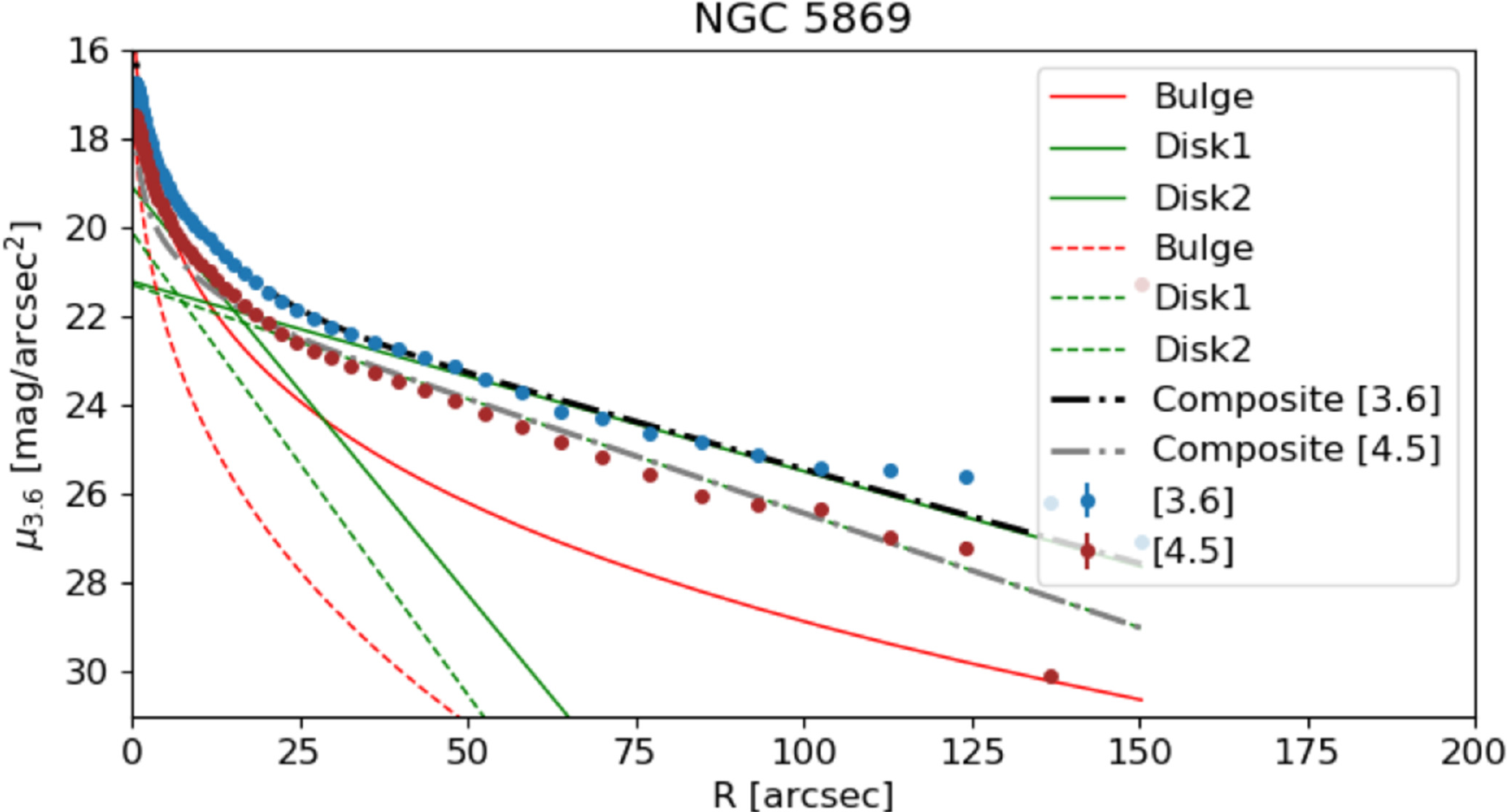}} \\
\subfloat{\includegraphics[width=0.48\textwidth]{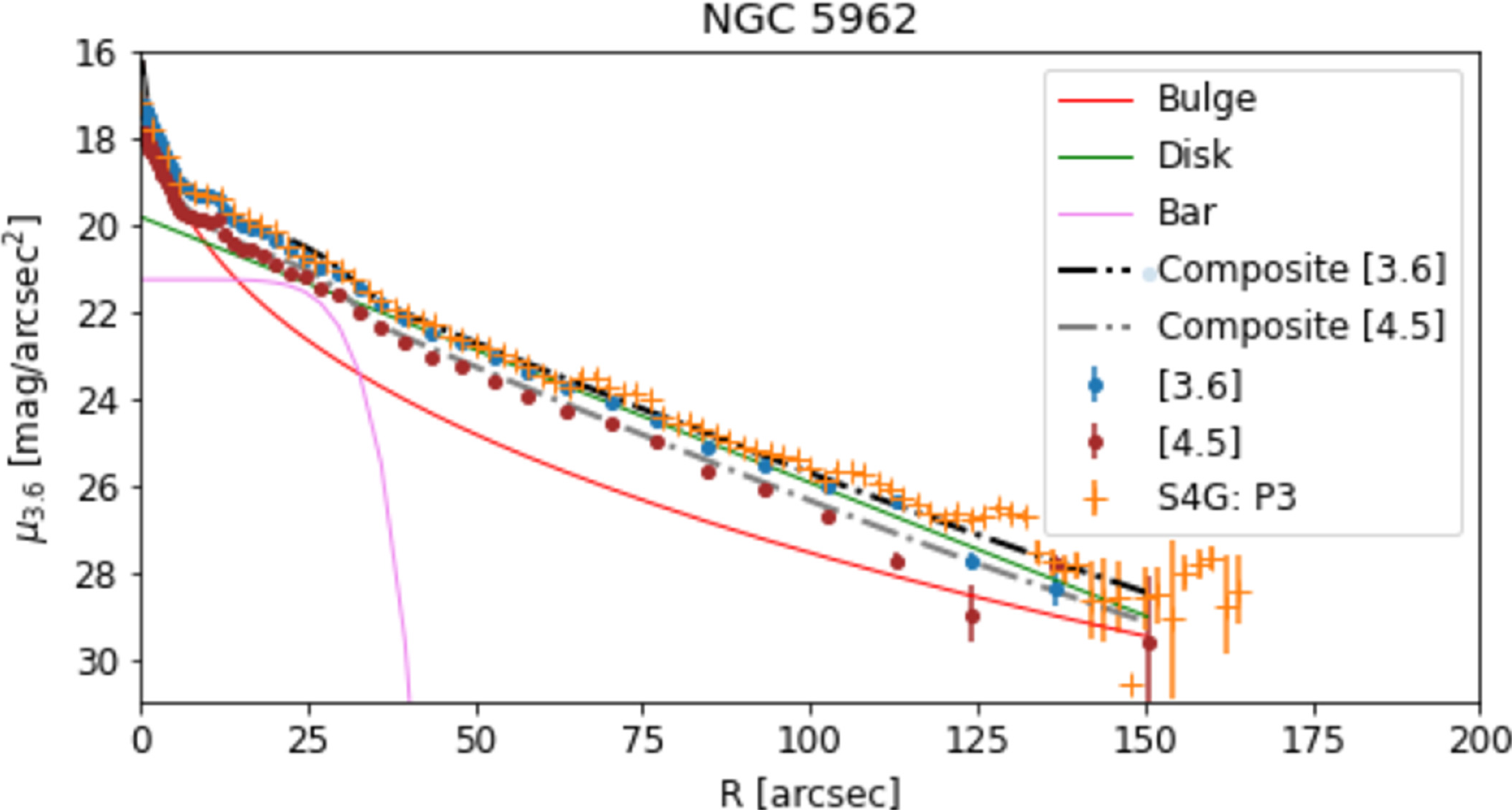}}
\subfloat{\includegraphics[width=0.48\textwidth]{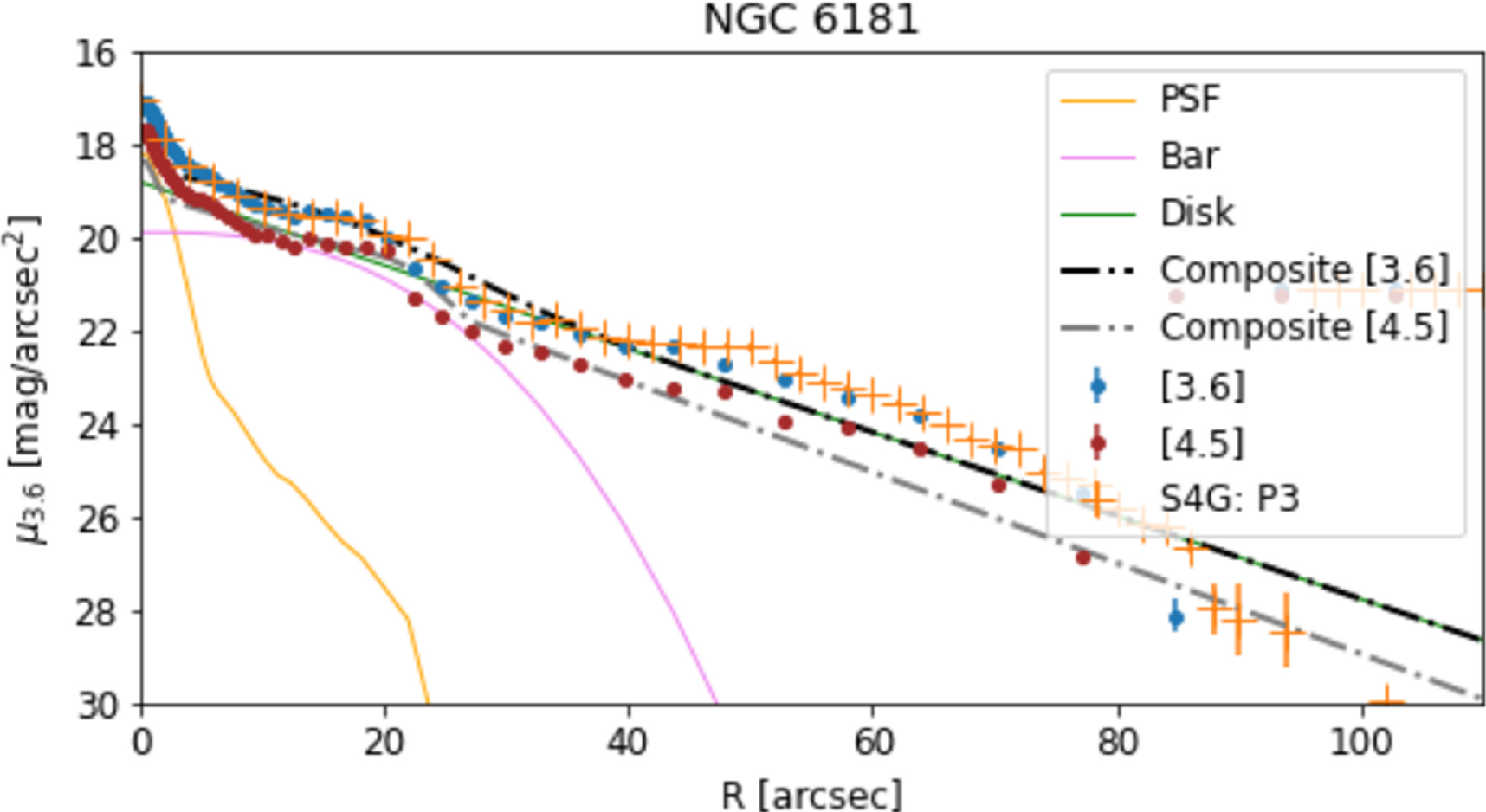}} \\
\subfloat{\includegraphics[width=0.48\textwidth]{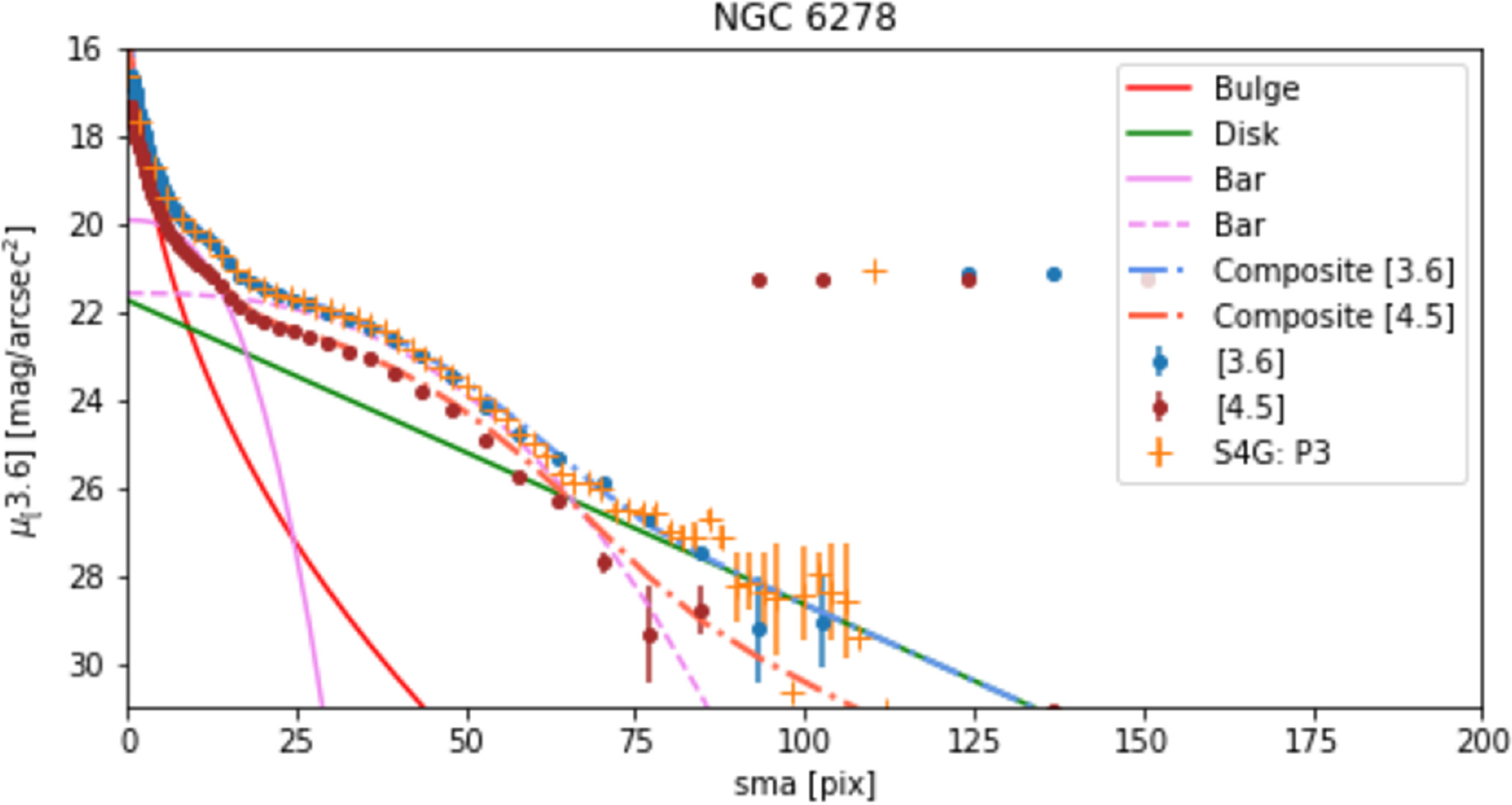}} 
\subfloat{\includegraphics[width=0.48\textwidth]{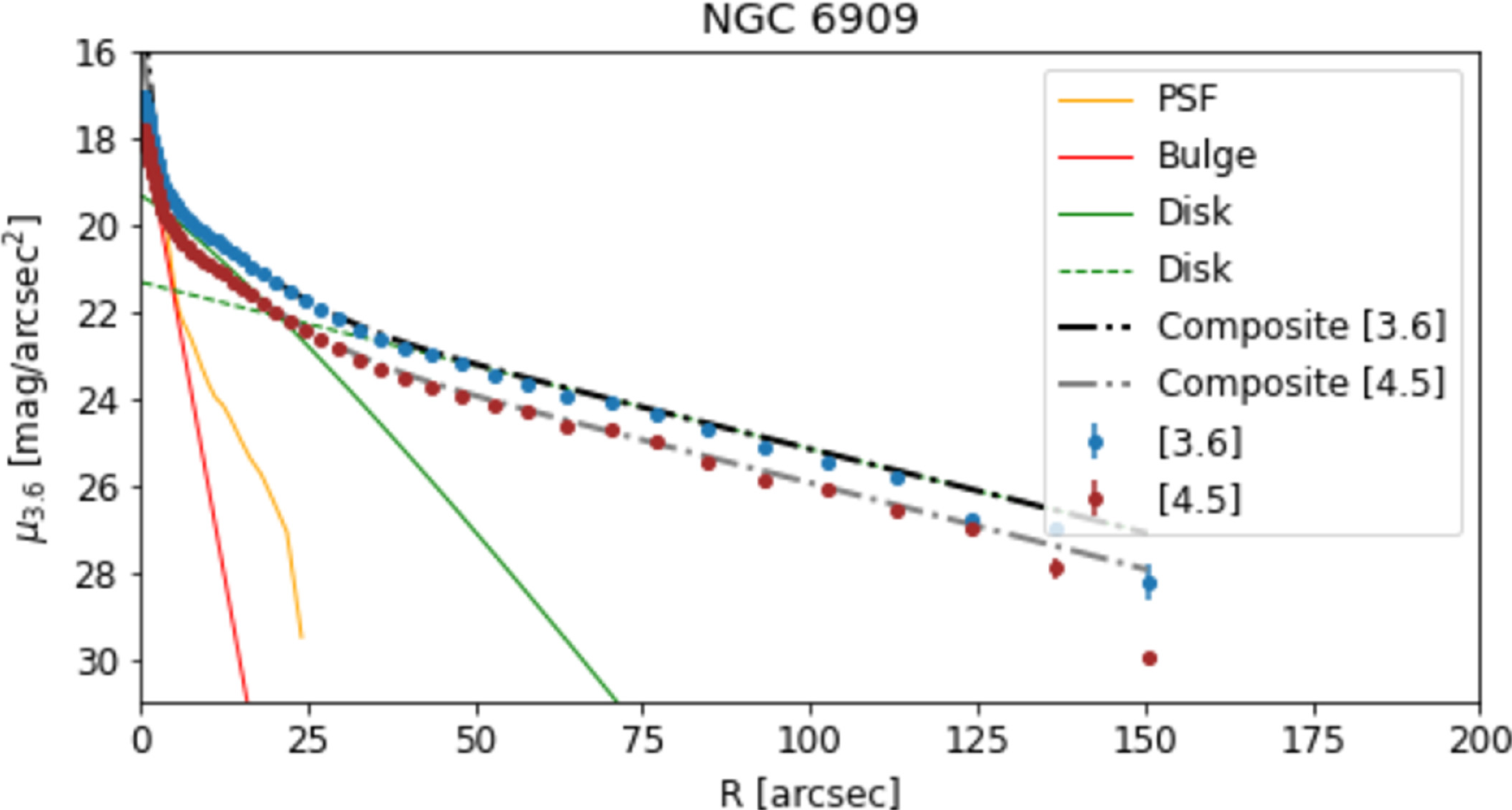}}\\

\caption[]{continued.}
\end{figure}
\begin{figure}
\ContinuedFloat 
\subfloat{\includegraphics[width=0.48\textwidth]{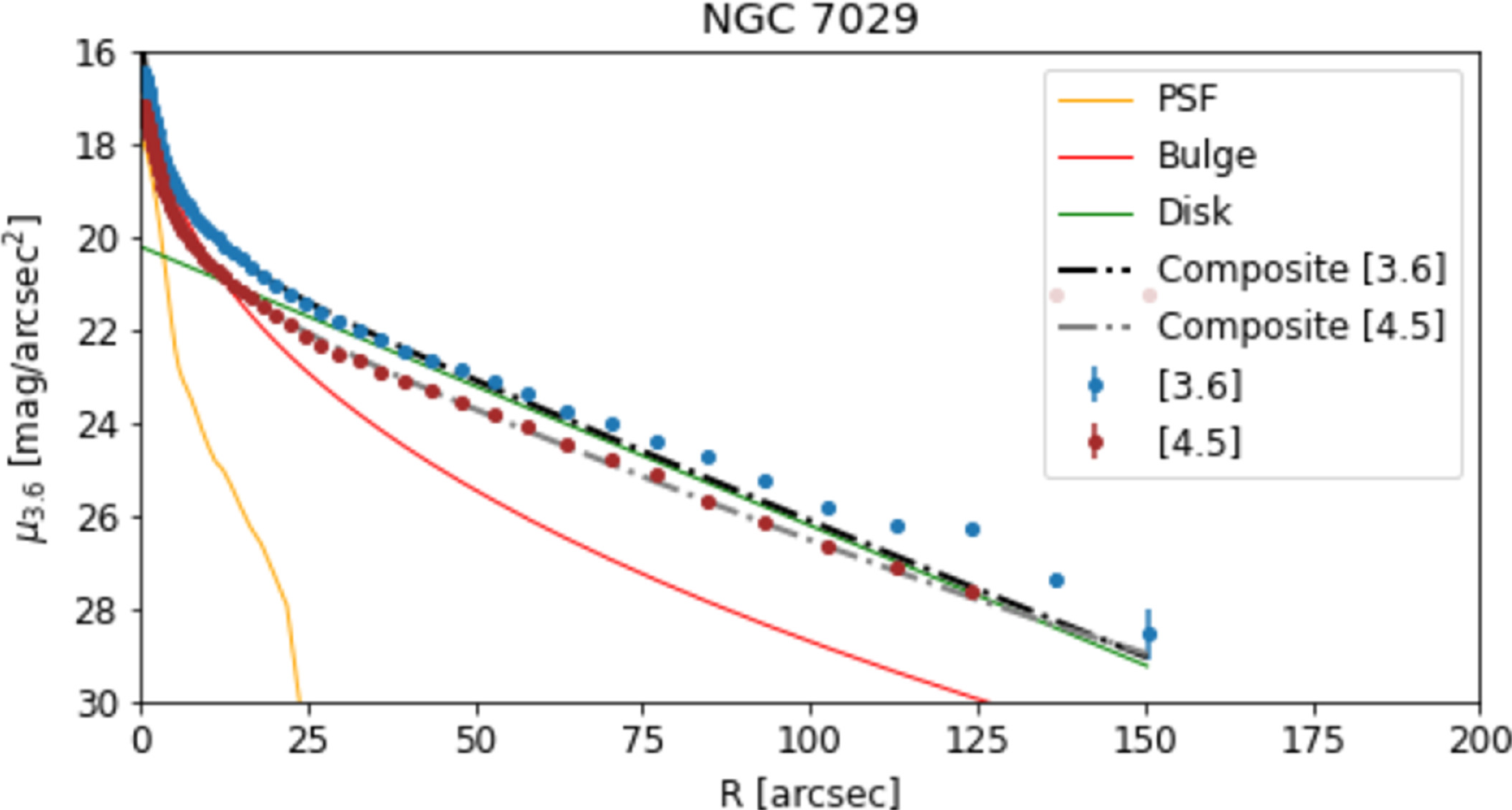}} 
\subfloat{\includegraphics[width=0.48\textwidth]{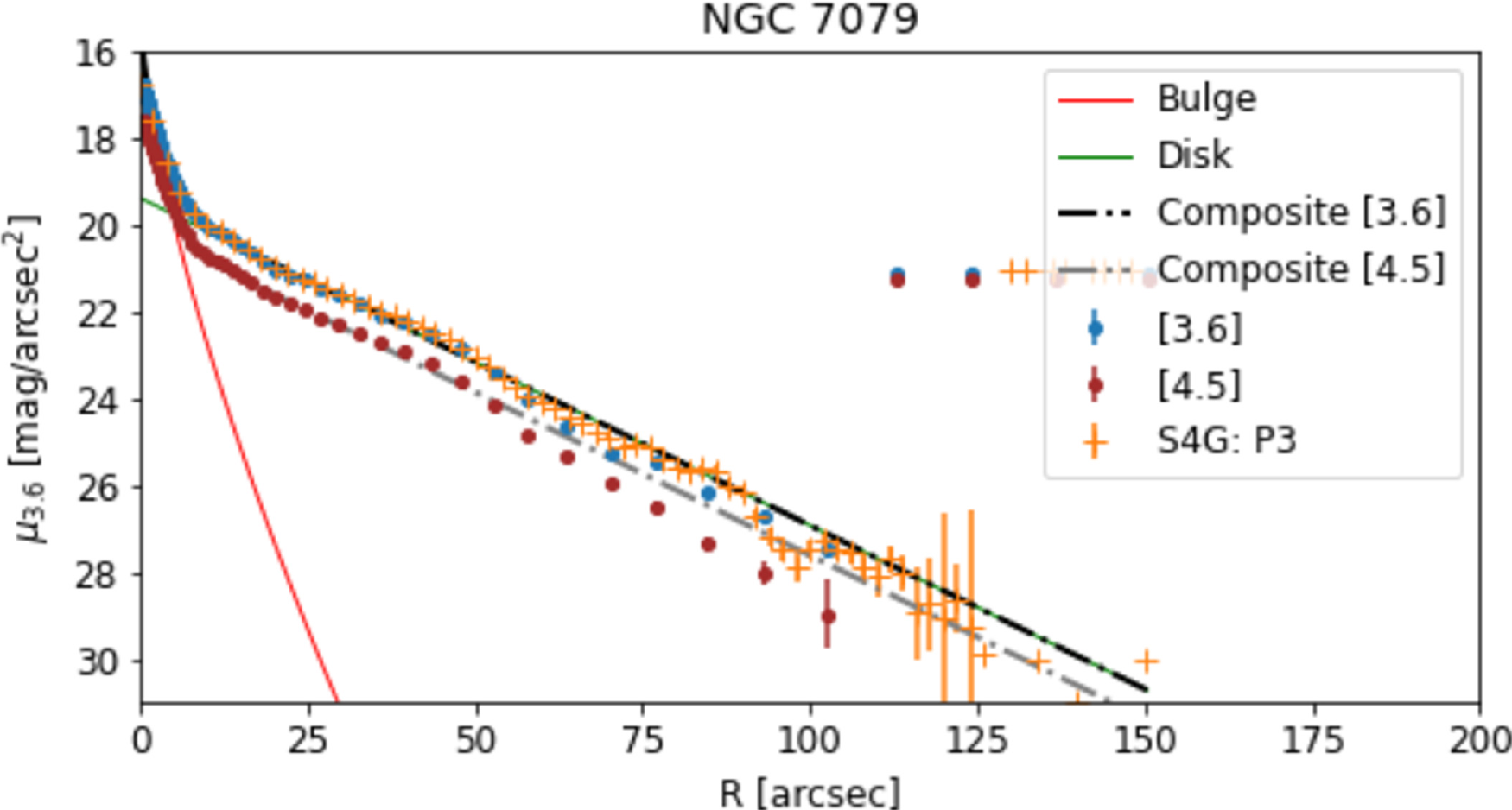}} \\
\subfloat{\includegraphics[width=0.48\textwidth]{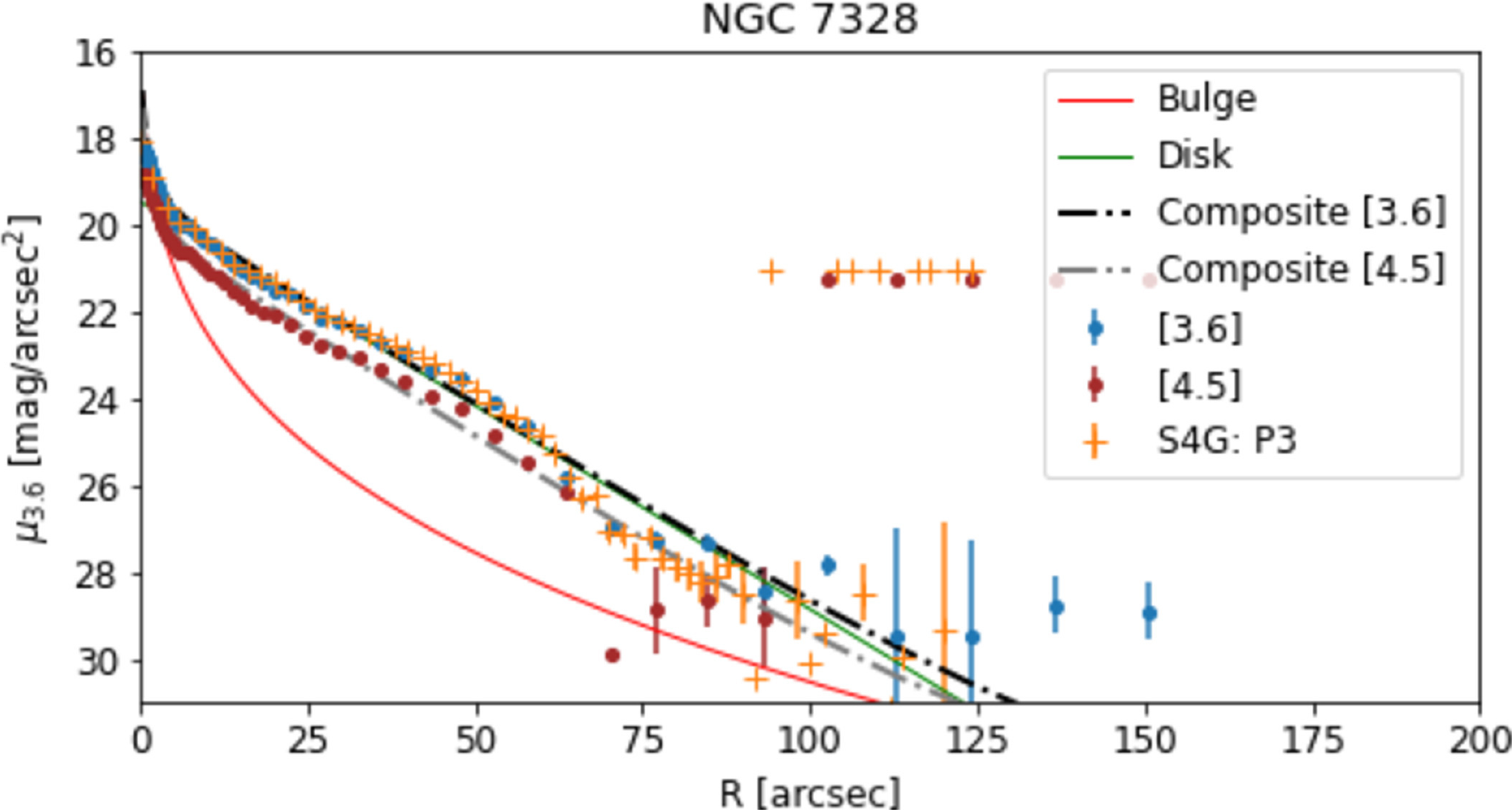}} 
\subfloat{\includegraphics[width=0.48\textwidth]{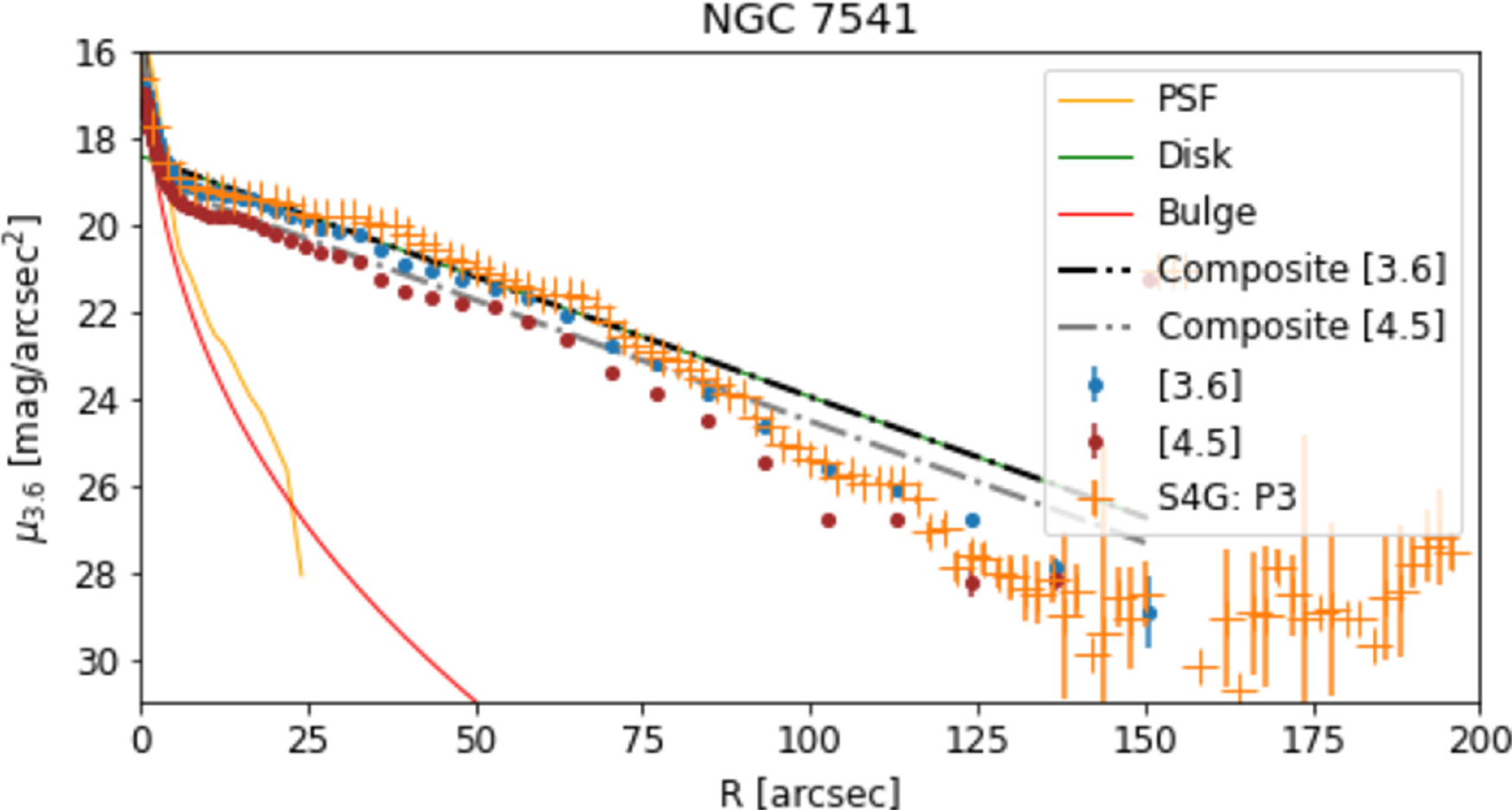}} \\
\subfloat{\includegraphics[width=0.48\textwidth]{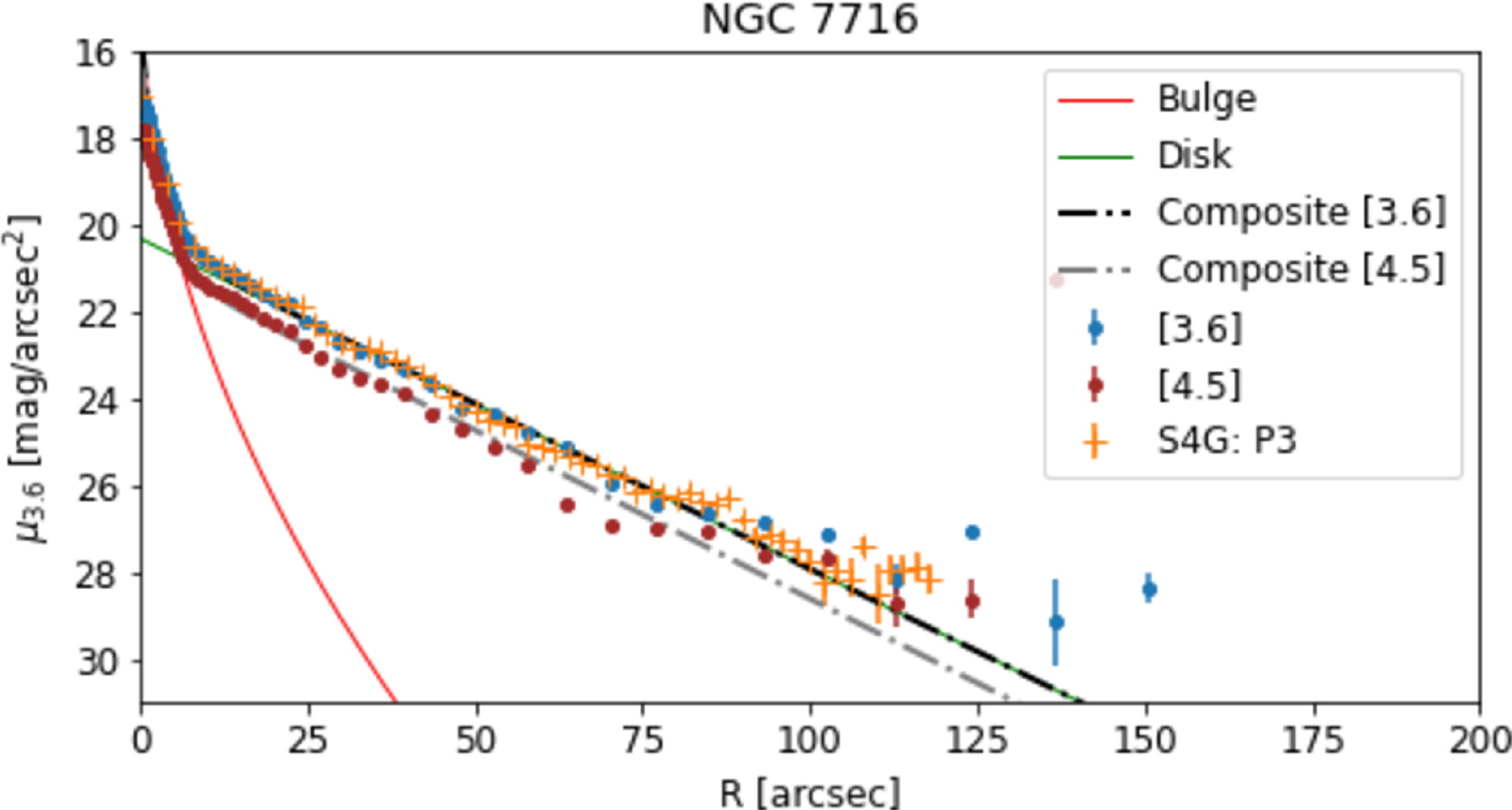}} 
\subfloat{\includegraphics[width=0.48\textwidth]{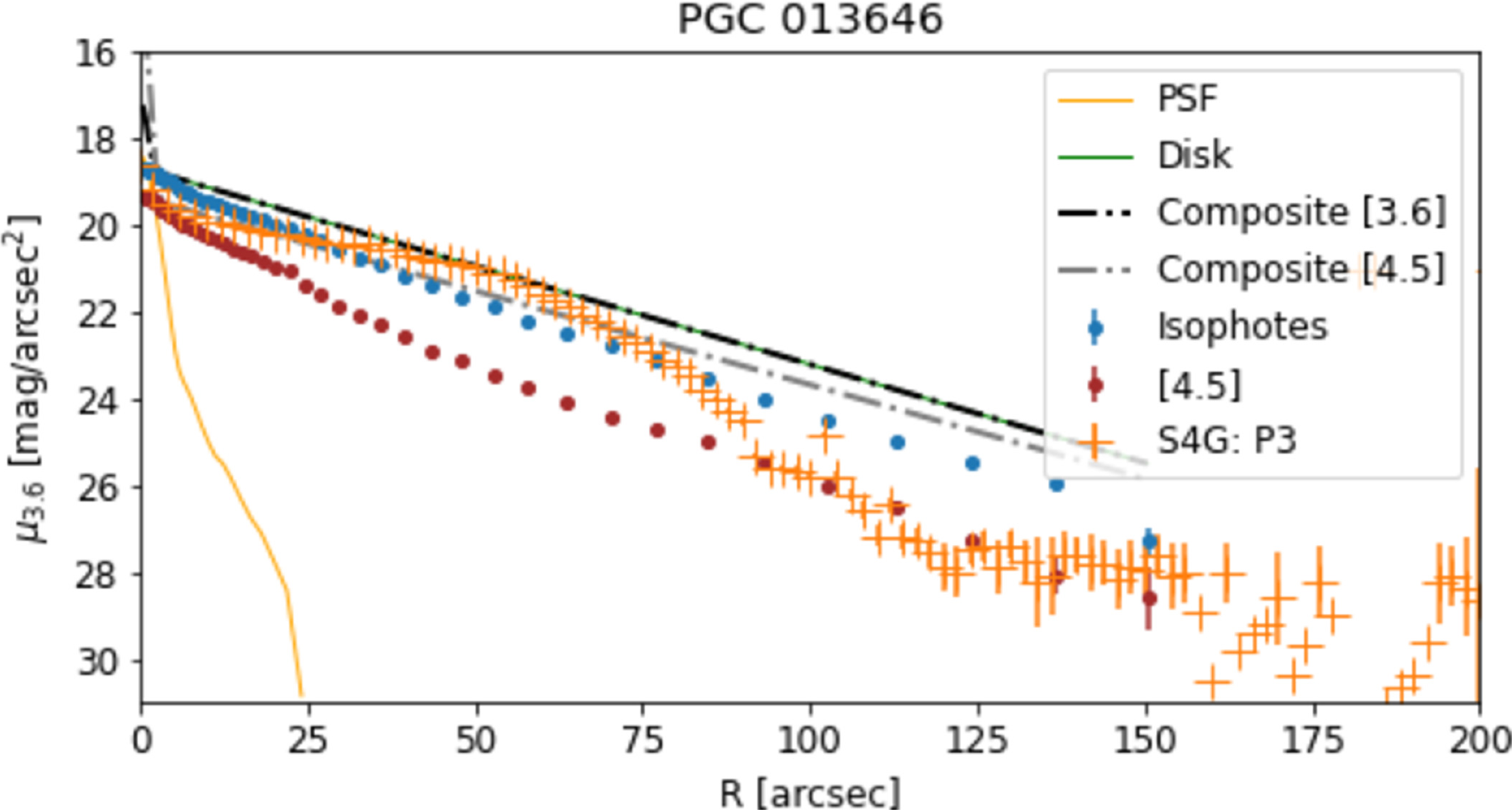}} \\
\subfloat{\includegraphics[width=0.48\textwidth]{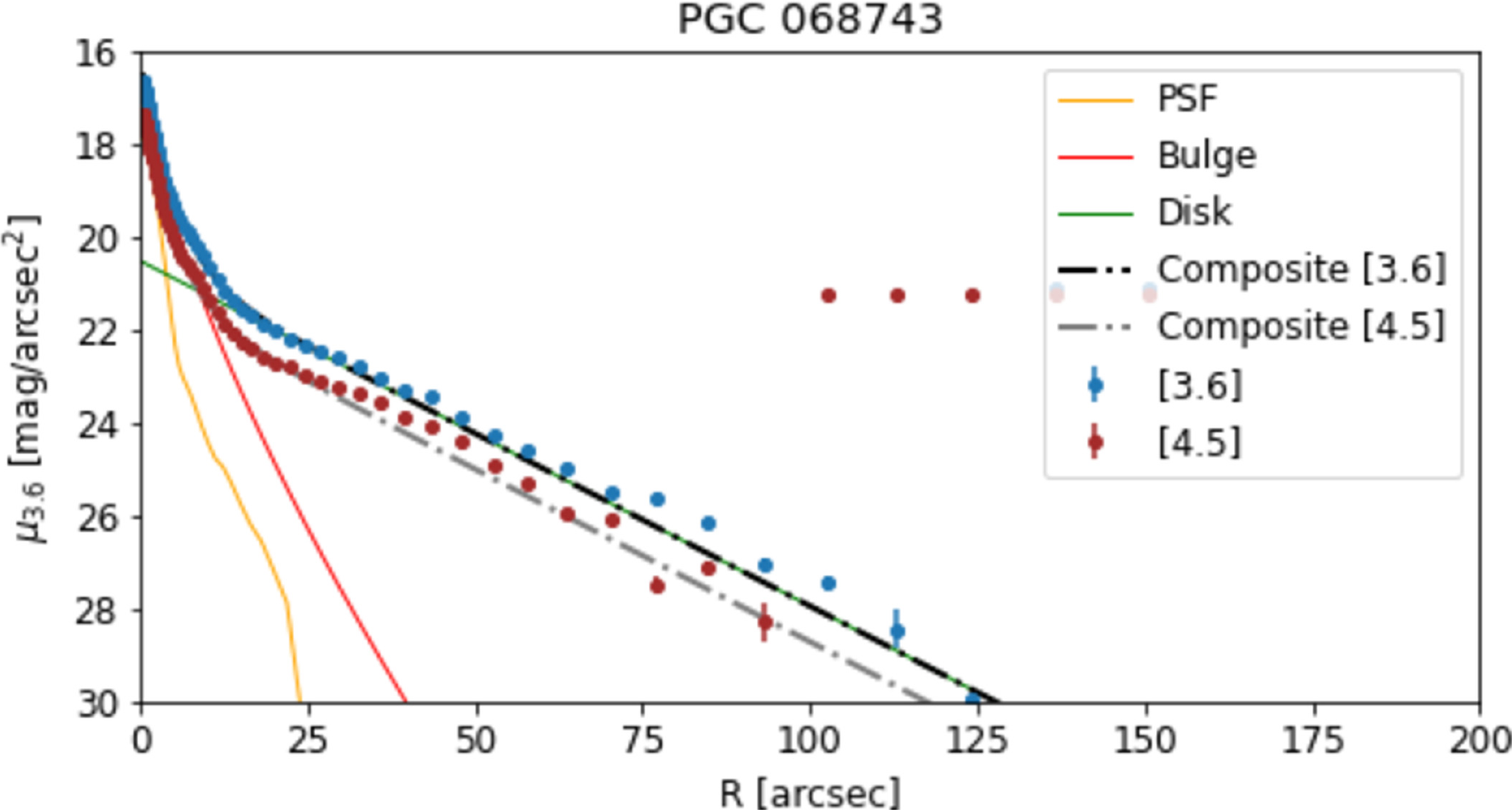}}
\subfloat{\includegraphics[width=0.48\textwidth]{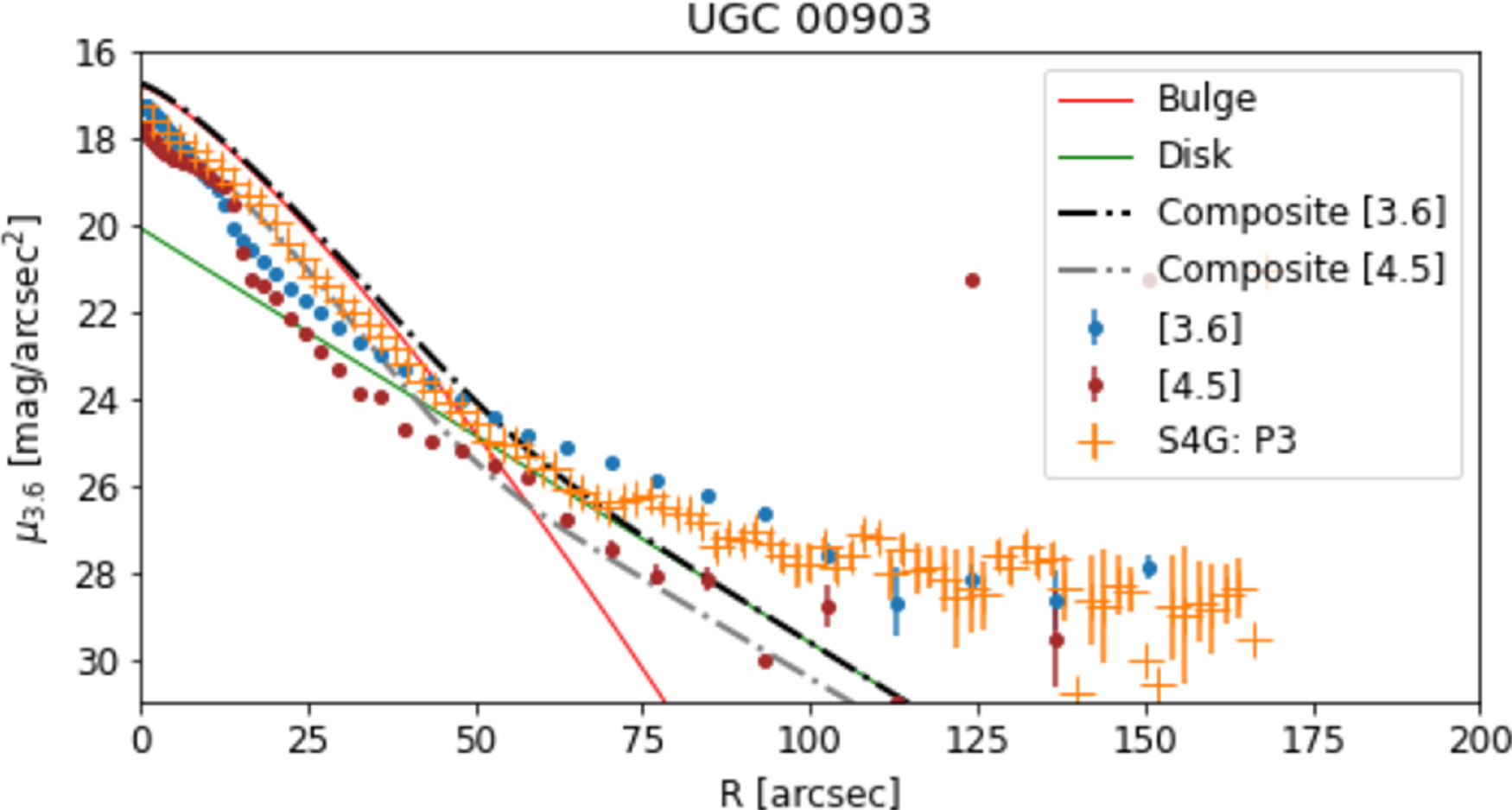}}\\
\caption[]{continued.}
\end{figure}

\end{appendix}

\end{document}